\documentclass[structabstract]{aa}  
\usepackage{graphicx}
\usepackage{txfonts}
\usepackage{natbib}

\begin{document}

\title{uvby-$\beta$ photometry of solar twins
\thanks{Based on observations 
collected at the H.L.~Johnson $1.5\,$m telescope at the Observatorio 
Astron\'omico Nacional at San Pedro M\'artir, Baja California, 
M\'exico.}
}
\subtitle{The solar colors, model atmospheres, and the \teff and metallicity scales}
\titlerunning{The colors of the Sun}

\newcommand{\teff}{$T_{\rm eff}$ }
\newcommand{\tsin}{$T_{\rm eff}$}

\author{
J. Mel\'endez\inst{1,2} \and
W. J. Schuster\inst{3} \and
J. S. Silva\inst{3} \and
I. Ram{\'{\i}}rez\inst{4} \and
L. Casagrande\inst{4} \and
P. Coelho\inst{5}
}


\institute{
Centro de Astrof\'{\i}sica da Universidade do Porto, Rua das Estrelas, 4150-762 Porto, Portugal.
\and
Departamento de Astronomia do IAG/USP, Universidade de S\~{a}o Paulo, Rua do Mat\~{a}o 1226, 
Cidade Universit\'{a}ria, 05508-900, S\~{a}o Paulo, SP, Brasil. 
$\; \; \;$ e-mail:  jorge@astro.iag.usp.br   \and
Observatorio Astron\'omico Nacional, Universidad Nacional Aut\'onoma de M\'exico,
Apartado Postal 877, Ensenada, B.C., CP 22800,\\
Mexico.$\;\;\;\;$   e-mail:  schuster@astrosen.unam.mx     \and
Max Planck Institute for Astrophysics, Postfach 1317, 85741 Garching, Germany. \and
N\'ucleo de Astrof\'{\i}sica Te\'orica, Universidade Cruzeiro do Sul, 
R. Galv\~ao Bueno, 868, Liberdade, 01506-000, S\~ao Paulo, Brasil.
}

\date{Received ...; accepted ...}

 
  \abstract
   {}
   {Solar colors have been determined on the $uvby$--$\beta$ photometric system
   to test absolute solar fluxes, to examine colors predicted by model atmospheres 
   as a function of stellar parameters (\tsin, log $g$, [Fe/H]), 
   and to probe zero-points of \teff and metallicity scales.
   }
   {New $uvby$--$\beta$ photometry is presented for 73 solar-twin candidates.
Most stars of our sample have also been observed spectroscopically to obtain
accurate stellar parameters. Using the stars that most closely resemble the Sun, 
and complementing our data with photometry available in the literature, the solar
colors on the $uvby$--$\beta$ system have been inferred.
Our solar colors are compared with synthetic solar colors computed from 
absolute solar spectra and from the latest Kurucz (ATLAS9) and MARCS model atmospheres.
The zero-points of different \teff and metallicity scales are verified
and corrections are proposed.
   }
   {Our solar colors are 
$(b-y)_{\odot}$ = 0.4105$\pm$0.0015, $m_{1, \odot}$ = 0.2122$\pm$0.0018,
$c_{1, \odot}$ = 0.3319$\pm$0.0054, and $\beta_{\odot}$ = 2.5915$\pm$0.0024.
The $(b-y)_{\odot}$ and $m_{1, \odot}$
colors obtained from absolute spectrophotometry of the Sun agree within
3-$\sigma$ with the solar colors derived here when the photometric 
zero-points are determined
from either the STIS HST observations of Vega or an ATLAS9 Vega model, but the
$c_{1, \odot}$ and $\beta_{\odot}$ synthetic colors inferred from absolute solar spectra
agree with our solar colors only when the zero-points based on the ATLAS9 model are
adopted.  The Kurucz solar model provides a better 
fit to our observations than the MARCS model.  For photometric values computed from the
Kurucz models, $(b-y)_{\odot}$ and $m_{1, \odot}$ are in excellent agreement with our 
solar colors independently of the adopted zero-points, but for $c_{1, \odot}$ and
$\beta_{\odot}$ agreement is found only when adopting the ATLAS9 zero-points.
The $c_{1, \odot}$ color computed from both the Kurucz and MARCS models is the most
discrepant, probably revealing problems either with the models or observations
in the $u$ band.  The \teff calibration of Alonso and collaborators has the poorest
performance ($\sim$140 K off), while the relation of Casagrande and collaborators
is the most accurate (within 10 K).  We confirm that the Ram\'irez \&
Mel\'endez $uvby$ metallicity calibration, recommended by
\'Arnad\'ottir and collaborators to obtain [Fe/H] in F, G, and K dwarfs, needs a small
($\sim$10\%) zero-point correction to place the stars and the Sun on the same
metallicity scale.  Finally, we confirm that the $c_1$ index in solar analogs has
a strong metallicity sensitivity.
}
   {}


\maketitle

%

\section{Introduction}

Photometry in the $uvby$--$\beta$ system \citep{str63,crawford66} 
is well suited to the determination of basic stellar atmospheric parameters
for F-, G-, and K-type stars through the color indices $(b$--$y)$, 
$m_{\rm 1} = (v$--$b) - (b$--$y)$ and $c_{\rm 1} = (u$--$v) - (v$--$b)$.
Several empirical calibrations exist in the literature to transform
$(b$--$y)$ or $\beta$ to $T_{\rm eff}$ (\citealp[e.g.][]{alo96,alo99,
cle04,ram05b}, hereafter RM05b; \citealp{hol07,cas10}), while
the $m_{\rm 1}$ index can be used to determine [Fe/H] in dwarfs 
(\citealp[e.g.][]{str64,gus72,ols84,sch89,mal94,hay02,mar02,mar04,ram05a}, 
hereafter RM05a; \citealp{twa07,hol07})
and giants \citep[e.g.][]{bon80,are96,hil00,ram04,cal07,cal09},
as reviewed by \cite{arn10}.
The evolutionary stage of stars can be determined using the
$c_{\rm 1}$ index together with other $uvby$ color 
indices \citep[e.g.][]{cra75,ols84,nis91,sch04,twa07}.

Since there are many difficulties in observing the Sun with the same instrumentation
as we observe other stars \citep[e.g.][]{ste57,gal64,cle79,tue82,loc92}, the Sun
cannot be used to set the zero-points of transformations between color indices 
and fundamental stellar parameters.
Accurate transformations are important in many areas of astrophysics. For example,
in the study of the primordial lithium abundance, an accurate \teff scale
is needed to compare the Li abundance of metal-poor stars with that obtained from
Big Bang Nucleosynthesis \citep[e.g.][]{mel10a,sbo10}.
The terrestrial planet signatures found in the chemical composition of the Sun
(\citealp{mel09}, hereafter M09; \citealp{ram09}, hereafter R09; \citealp{ram10}) shows that for accurate comparisons between the Sun and stars
an accurate temperature  scale must be used in the determination of chemical abundances. 
Although there is inconclusive evidence about whether the Sun is too metal-rich
with respect to stars of similar age and Galactic orbit \citep{hay08,hol09} implying
that the Sun could have been born in the inner part of the Galaxy \citep{wie96}, this
apparent offset between the Sun and stars could be partly due to zero-point errors in
the photometric metallicity scale (see e.g. Table 3 of \citealt{arn10} for
systematic differences between spectroscopic and photometric metallicities).
Furthermore, we note that zero-point errors in the \teff scale, as well as errors
in the metallicity scale, would introduce systematic errors in the ages
of stars determined from isochrones.

According to their similarity to the Sun, 
stars can be classified as ``solar-type stars'' (late F to early K stars),
``solar analogs'' (G0-G5 dwarfs with solar metallicity within $\sim$ a factor of 2-3),
and ``solar twins'' (stars almost identical to the Sun) (\citealp[e.g.][]{secchi1868,cay96,
sod98}; M09).  Many works have used either solar-type stars or
solar analogs to infer solar colors to improve or check the 
effective temperature scale, the performance of model atmospheres, and 
the absolute flux calibration of the Sun
(\citealp[e.g.][]{pet28,kui38,ste45,ste57,joh62,kro63,van65,
cro72,ols76,sch76,bar78,har78,har80a,har80b,cle79,
cay81,nec81,chm81,tue_sch82,mag83,tay84,mit85,nec86,van89,
gra92,gra95,str94,tay94,cay96,col96}, hereafter C96; 
\citealp{hau96,bes98,mir98,sek00,mel03}; RM05b; \citealp{hol06,pas08,rie08,cas10}).

Solar twins have spectra almost identical to the Sun \citep{cay96}, hence
are better suited to setting the zero-points of fundamental calibrations, especially
since they are found without assuming {\em a priori} a temperature scale,
but their identification is based purely on a model-independent analysis
of their spectra with respect to a solar spectrum obtained with the same
instrumentation.  However, until only a few years ago one solar twin was known
\citep[18 Sco;][]{por97,sou04}. The situation has changed dramatically 
with the identification of many stars similar to the Sun
(\citealp{mel06}, hereafter M06; \citealp{mel07}, hereafter MR07;
\citealp{tak07}, hereafter T07; \citealp{tak09}, hereafter T09;
M09; R09; Baumann et al. 2010, hereafter B10).
It is now feasible to
use solar twins with accurately determined stellar parameters to test
the predictions of model atmospheres and the accuracy of empirical 
photometric calibrations. In the present work, we perform this study for
the widely used $uvby$--$\beta$ system.

\section{Photometric observations}

\subsection{Selection of the sample}

Before starting our systematic survey of stars
similar to the Sun in the Hipparcos catalogue, we performed
pilot observations of solar-twin candidates to test our 
selection criteria. Our pilot study (M06) found 
zero-point offsets in our \teff scale (RM05b).
We therefore applied small corrections to the solar colors predicted by RM05b
to increase our chances of finding stars resembling the Sun. 
In particular, we searched for solar twins around the 
Tycho color $(B-V)_T$ = 0.7225 instead of $(B-V)_T$=0.689 
predicted by our earlier color-\teff relations (RM05b).
Our improved \teff scale
(Casagrande et al. 2010) indeed corrected the zero-point problem,
and now predicts a solar $(B-V)_T$=0.730, in good agreement
with our tentative correction.

The main parameters used to select solar-twin candidates from the Hipparcos
catalogue were parallaxes and $(B-V)_T$ colors. Additional criteria
used (when available) were other optical-infrared colors (e.g. $V_T-K$, $b-y$),
photometric variability, information on multiplicity, previous
literature values for [Fe/H] \citep[obtained from an updated version of the] 
[catalogue]{cay01}, rotation, and chromospheric activity.
We found initially about one hundred stars satisfying our selection criteria
within 75 pcs. We later expanded our search to cover 
the whole Hipparcos catalogue, increasing our sample by about 1/3.

\subsection{Observations and data reduction}

The $uvby$--$\beta$ data presented here in Table 1 for the solar twins were taken
using the H.~L. Johnson $1.5\,$m telescope at the San Pedro M\'artir Observatory,
Baja California, M\'exico (hereafter SPM), and the same six-channel $uvby$--$\beta$
photoelectric photometer as for the northern observations of Schuster \& Nissen
(1988, hereafter SN), for all the $uvby$--$\beta$ observations of Schuster et al.
(1993, hereafter SPC), the northern data of very-metal-poor stars
by Schuster et al.~(1996), the $uvby$--$\beta$ data for very-metal-poor stars
in Table 1 of Schuster et al.~(2004), and the $uvby$--$\beta$ data for high-velocity
and metal-poor stars in Table 1 of Schuster et al.~(2006).  The new $uvby$--$\beta$
values for solar twins included here in Table 1 were taken during three observing
runs in November 2007 (8 nights), April 2008 (7 nights), and September 2008 (4 nights).

The $uvby$--$\beta$ solar-twin data 
were taken and reduced using techniques very nearly the same as for SN and SPC
(see these previous papers for more details). The four-channel $uvby$ section
of the SPM photometer is really a spectrograph-photometer that employs exit
slots and optical interference filters to define the bandpasses.  The grating
angle of this spectrograph-photometer was calibrated using a cadmium lamp at
the beginning of each observing run to position the spectra on the exit slots
to within about $\pm 1{\AA}$.  Whenever possible, extinction-star observations
were made nightly over an air-mass range of at least 0.8 (see Schuster \&
Parrao 2001; also Schuster et al.~2002), and spaced throughout each night
several ``drift'' stars were observed symmetrically with respect to the local
meridian (two hours east and then two hours west).  Using these observations,
the atmospheric extinction coefficients and time dependences of the night
corrections could be obtained for each of the nights of observation
(see Gr{\o}nbech et al.~1976).  Finding charts were employed at SPM to
confirm identifications of the program stars and to select regions for the
``sky'' measurements.  As for previous studies, such as SN and SPC, the program
stars were observed at SPM to at least 50 000 counts in all four channels of
$uvby$ and to at least 30 000 counts for the two channels centered at H$\beta$.
For all program stars, the sky background was measured until its contributing
error was equal to or smaller than the error in the stellar count.  At SPM, an
attempt was made to obtain three or more independent $uvby$ observations for
each of the program stars, i.e. photometric observations during at least three
independent nights; this aim was achieved, or exceeded, for all solar twins
except HIP75923 and HIP77883 for which we obtained two observations each.

As for the SN and SPC catalogues, all data reduction was carried
out following the precepts of Gr{\o}nbech et al.~(1976) using computer 
programs kindly loaned by T. Andersen (see Parrao et al.~1988).  At SPM, the
$uvby$--$\beta$ standard stars observed were taken from the same lists as
for the previous catalogues, and are mostly secondary standards from
the catalogues of Olsen (1983, 1984).  The reduction programs
create a single instrumental photometric system for each observing run,
including nightly atmospheric extinctions and night corrections with
linear time dependences.  Then transformation equations from the
instrumental to the standard systems of $V$, $(b$--$y)$, $m_{\rm 1}$,
$c_{\rm 1}$, and $\beta$ are obtained using all standard stars observed
during that observing period.  The equations for the transformation to
the standard $uvby$--$\beta$ system are the linear ones of Crawford \&
Barnes (1970) and Crawford \& Mander (1966).  Small linear terms in
$(b$--$y)$ are included in the standard transformation equations for
$m_{\rm 1}$ and $c_{\rm 1}$ to correct for bandwidth effects in the $v$
filter.  Our $y$ measurements were transformed onto the V system of Johnson
et al.~(1966).

Thirty-six $uvby$--$\beta$ standard stars were employed during the
observing run of November 2007 providing instrumental photometric errors
ranging from 0.002 mag. in $(b$-$y)$ to 0.009 mag. in $c_{\rm 1}$, and errors in the
transformations to the standard photometric system from 0.004 to 0.011
magnitudes, respectively.  For April 2008, these values were for 33
standard stars, 0.002--0.010 mag., and 0.005--0.013 mag., respectively, and
for September 2008, for 35 standard stars, 0.002--0.008 mag., and 0.006--0.009
mag., respectively.  Instrumental and transformation errors in the magnitude
V and in the indices $m_{\rm 1}$ and $\beta$ were always intermediate in value between
those given above.

\subsection{The catalogue}

Table 1 presents the $uvby$--$\beta$ catalogue for the 73 solar-twin candidates 
observed at SPM. Column 1 lists the Hipparcos number; 
Col.~2 gives the HD  (or BD) number, Col.~3 the $V$ magnitude on the standard
Johnson $UBV$ system; and Cols.~4--6 and 9, the 
indices $(b$--$y)$, $m_{\rm 1}$, $c_{\rm 1}$ and $\beta$ on the
standard systems of Olsen (1983, 1984), which are essentially the systems
of Crawford \& Barnes (1970) and Crawford \& Mander (1966) with north-south
systematic differences corrected.  Columns 7, 8, and 10 give $N_V$, $N_{uvby}$,
and $N_{\beta}$, the total numbers of independent V, $uvby$, and $\beta$
observations.

A very small subset of our photometric observations was made through light
cirrus clouds in the absence of moonlight.  It has been well documented
(e.g. SN; Olsen, 1983) that observations in the indices $b$--$y$, $m_{\rm 1}$,
$c_{\rm 1}$, and $\beta$ made with simultaneous multichannel photometers are
not affected in any significant way by light (or even moderate) cirrus, while
the V magnitude, obtained from only the $y$ band, is affected.  For this
reason, a few of the solar twins, such as HIP60314, HIP74389, and HIP118159,
have fewer independent observations of the V magnitude than the indices.

\subsection{Comparisons with other photometric data}
We searched for $uvby$-$\beta$ photometry in the General Catalogue of Photometric
Data \citep{mer97,hau98} and found data for 
23 solar-twin candidates and 12 solar analogs
(observed by ourselves spectroscopically for
other projects). 
The photometry of these 
35 stars is given in Table 2.

The accuracy of $uvby$-$\beta$ photometry obtained at SPM has
been extensively tested (e.g. SN; \citealp{are90,sch93,sch96}),
and we illustrate below that our solar-twin photometry is
also in excellent agreement with the literature, in all cases 
with mean differences well below 0.01 magnitudes.
There are 12 stars in common between our sample and previous work.
The average difference (ours - literature) in $(b$--$y)$ is only
$-0.001$ ($\sigma$ = 0.004). Our V Johnson photometry is also in
good agreement, with a mean difference (ours - literature)
of only +0.001 ($\sigma$ = 0.006). The colors $m_1$ and $c_1$
also compare well, with a difference (ours - literature) of
only +0.002 ($\sigma$ = 0.006) and +0.006 ($\sigma$ = 0.015).

Considering the excellent agreement between the photometry available
in the literature and in our own data sets, and considering the
previously shown accuracy and precision of the SPM $uvby$-$\beta$
photometry, we conclude that both data sets, of Tables 1 and 2, provide
solar-color indices very close to the standard $uvby$-$\beta$ system.

\section{The solar-color indices}
Our observations comprise the largest photometric data set yet taken 
of stars very similar to the Sun in the $uvby$--$\beta$ system.
Adding other photometry available in the literature and
using our own accurate stellar parameters 
(M09; R09; \citealp{mel10b}, hereafter M10b), and taking into account
the variations in colors with \tsin, log $g$ and [Fe/H], we can
infer the ``solar'' colors by interpolating them to the stellar
parameters of the Sun: \teff = 5777 K, log $g$ = 4.44, and [Fe/H] = 0.00
(\citealp[e.g.][]{cox00,gra05}; RM05b).

The quality of our stellar parameters (M06; MR07; M09; R09; M10b)
is very high because both the Sun (reflected light of asteroids) and the 
solar twins were observed with the same instrumentation during the
same observing runs, and all data reduction and analysis were performed
in an identical way. Our spectra have typically a resolution of
60,000 and S/N$\sim$200 for stars observed with the 2.7m telescope at McDonald 
and S/N$\sim$450 for stars observed with 6.5m Magellan Clay telescope at
Las Campanas. Errors as low as $\sim$25 K in \tsin, 0.04 dex in log $g$,
and 0.025 dex in [Fe/H] can be obtained in the best cases, and abundance
ratios with errors as low as 0.01-0.02 dex have been obtained with the above
data, showing that the Sun is a star with a peculiar chemical composition
(M09; R09).
Additional spectra available in the literature for solar-twin candidates 
were analyzed by B10 using similar techniques.  The spectra are from 
HARPS observations available at the ESO archive and from the 
S$^4$N database \citep{all04}\footnote{S$^4$N: Spectroscopic Survey of Stars in the Solar
Neighborhood; available online at http://hebe.as.utexas.edu/s4n/.}. In both cases,
a solar spectrum 
taken with the same instrumentation was employed in the differential analysis. 

The stellar parameters were determined homogeneously by our own group 
(M06, MR07, M09, R09, M10b, B10) using Kurucz model atmospheres and a 
line-by-line differential analysis with respect to the Sun.  The \tsin,
log $g$, [Fe/H], and microturbulence were determined iteratively until
both the differential excitation equilibrium of FeI lines and the
differential ionization balance of FeI and FeII, were achieved.  The
microturbulence was also determined simultaneously, by requiring 
no dependence of the iron abundance (from FeI lines) on the reduced
equivalent width.

Since our solar-twin sample spans a relatively narrow range in 
atmospheric parameters relative to the Sun, 
in principle simple
linear fits of color versus (vs.) each parameter (\tsin, log $g$, and [Fe/H])
would provide sufficiently good estimates of the solar colors.
As suggested by the referee, a global fit \citep[e.g.][]{mit85}
to all stellar parameters (color = $f$(\tsin, log $g$, [Fe/H]))
would be preferable due to the mutual interdependence of the
stellar parameters. Fortunately our sample includes also
solar analogs covering a broader range in colors and 
stellar parameters than the solar twins, so that the
dependence on the different stellar parameters can be
well determined by a global fit. The following formula was employed:

\begin{equation}
color = A + B \, (T_{\rm eff} - 5777) \, + \, C \, ({\rm log} g - 4.44) + D \, {\rm [Fe/H].}
\end{equation}

\noindent The advantage of this equation is that $A$ will give us
directly the solar color, while its uncertainty could be 
determined from the error in $A$ or from the scatter of the fit.

From our sample of stars, we selected a group of solar twins
with \tsin, log $g$, and [Fe/H] within 100 K, 0.1 dex, and 0.1 dex
of the solar parameters\footnote{Strictly speaking, a solar-twin star 
must be identical to the Sun within the observational uncertainties.
Our definition based on derived stellar parameters is simply more practical
because it is less dependent on the quality of 
the stellar and solar spectra available.} given above. 
To perform a more robust global fit, we extended our solar twin 
sample with solar analogs covering a range in \tsin, log $g$, and [Fe/H].
Thus, to explore the metallicity dependence of the colors, we
selected a sample of solar analogs with the same constraints in
\teff and log~$g$, but covering a broader range in
metallicity ($-0.4 <$ [Fe/H] $< -$0.1 dex, +0.1 $<$ [Fe/H] $<$ +0.4 dex).
In a similar way, to improve the fit to \tsin, we
selected a group of solar analogs with the same
constraints on log~$g$ and [Fe/H] as the solar twins,
but with \teff in the range ($5617 <$ \teff $< 5677$ K, $5877 <$ \teff $< 5937$ K).
Finally, to fit the trend with log $g$,
we used a sample of solar analogs with the same constraints 
on \teff and [Fe/H] as the solar twins, but covering a
broader range in log $g$ ($4.29 \leq$ log $g < 4.34$ dex,
$4.54 \leq$ log $g < 4.59$ dex).

The stellar parameters were taken from our work on solar twins
(M06; MR07; M09; R09; M10b; B10)
and complemented in some cases with 
other accurate values available in the literature 
(\citealp{val05}, hereafter VF05; \citealp{luc06}, hereafter LH06;
T07; \citealp{sou08}, hereafter S08; T09). The adopted stellar parameters
for the solar twins and solar analogs are given in Table 3.

The global fits to $(b$--$y)$ had only three outliers, HIP 7245,
HIP 81512, and HIP 88427,
which seem too red in $(b$--$y)$ for their \teff 
and [Fe/H].
Using the \cite{kar10} intrinsic-color calibration, we find that these 
three stars may be
slightly reddened (E(b-y) $\sim$ 0.015 - 0.023), 
although we have to bear in mind
that the accuracy of the reddening calibration is of the same order.
\footnote{The star HIP 79186 also seems reddened according to the
\cite{kar10} calibration, but the stellar parameters of this star
(\tsin, log $g$, and [Fe/H]) = (5709 K, 4.27 dex, -0.12 dex) (R09)
do not fall either in our solar-twin or our solar-analog samples, 
so it was not considered in the global fits.}
These stars were removed from the 
global fits.
The results from the fits are presented in Table 4 and in Figs.~1--4,
where
filled circles represent the solar twins and open circles the solar analogs across a
broader range of stellar parameters.

\subsection{The $(b$--$y)$ solar color}

\begin{figure}
\resizebox{\hsize}{!}{\includegraphics{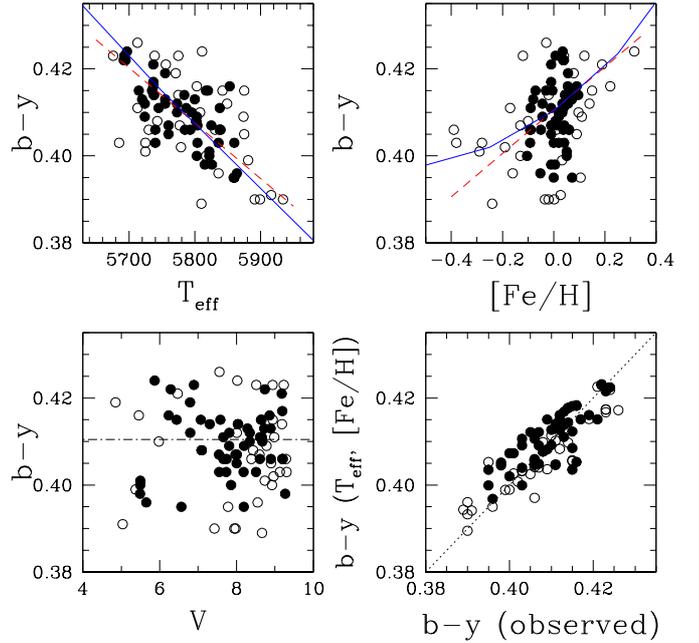}}
\caption{$(b$--$y)$ vs.\ \teff (upper left panel), [Fe/H] (upper right panel),
and V magnitude (lower left panel, with a dot-dashed line at $(b-y)_{\odot}$).
The results of the global fit vs.\ the observed $(b-y)$ color is presented in
the lower right panel, with the dotted line indicating equality.  Solar twins
and solar analogs are represented by filled and open circles, respectively. 
The dependences of the fit on \teff and [Fe/H] are shown by
dashed lines, while the relative predictions of MARCS models 
(normalized to our inferred solar colors) are shown by solid lines.
}
\label{bysyn}
\end{figure}

The global fit of $(b$--$y)$ (Fig.~1 and Table 4) shows
strong dependences on \teff (at the 14-$\sigma$ level)
and [Fe/H] (10-$\sigma$). There is no dependence (within the errors)
on log $g$, therefore this parameter was excluded from the global fit.
The star-to-star scatter
from the fit is only 0.005 magnitudes, which is
what is expected from the observational
uncertainties (0.004-0.006 magnitudes).

The standard error (s.e.) in the solar color was obtained from the
observed star-to-star scatter and the number of data-points
(s.e. = $\sigma/\sqrt{\rm sample \, size}$). The error in  
$(b$--$y)_\odot$ (and the other solar colors) was
conservatively adopted as three times the standard error. Thus, we propose

\begin{equation}
(b-y)_{\odot} = 0.4105 \; (\pm 0.0015).
\end{equation}

A plot of $(b$--$y)$ vs.\ V magnitude can help us to
reveal whether fainter (i.e., more distant) stars
bias the above derived solar color, due to
possible interstellar reddening. This plot
is shown in the lower left panel of Fig. 1. As can be
seen, there is no trend with V magnitude. 
A linear fit of $(b$--$y)$ vs.\ V, indeed shows a
zero slope within the errors (slope = 0.0003 $\pm$ 0.0008).
We note that most stars of the sample used in the global fit 
are brighter than V = 9, i.e.\ closer than $\sim$ 68 pc.
Even the faintest stars (V $\sim$ 9.3)
extend only to $\sim$78 pc.
According to NaI interstelar absorption maps, very little
NaI absorption is detected for distances up to $\sim$
80pc from the Sun \citep{lal03,wel10}.
Thus, most of our sample is not expected to be significantly affected by
interstellar absorption. As already mentioned above,
the few stars that show some small sign of interstellar
reddening were not included in the global fits.

\subsection{The $\beta$ solar color}

The global fit of $\beta$ shows a dependence on
\tsin, log $g$, and [Fe/H] (Table 4). Interestingly, the
strongest dependence is with [Fe/H] (slope
significant at the 5-$\sigma$ level),
while the dependence with both 
\teff (2.9 $\sigma$) and log $g$ (2.7 $\sigma$)
is only at the 3-$\sigma$ level.
In Fig.~2, we show the dependence of the $\beta$ color
index with respect to \teff and [Fe/H]. The scatter from
the fit is 0.006 mags, which is compatible with the observational errors in $\beta$.

The adopted $\beta$ solar color is

\begin{equation}
\beta_{\odot} = 2.5915 \; (\pm 0.0024).
\end{equation}

\begin{figure}
\resizebox{\hsize}{!}{\includegraphics{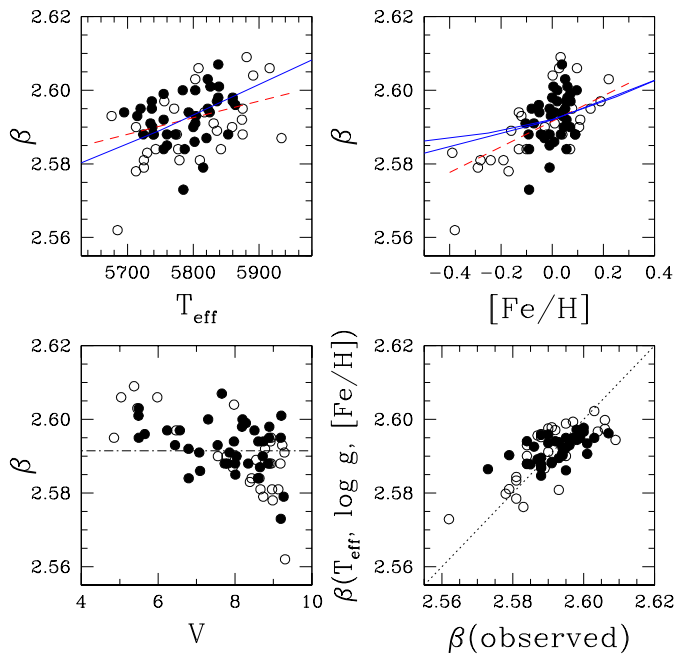}}
\caption{
$\beta$ vs.\ \teff (upper left panel), [Fe/H] (upper right panel),
and V magnitude (lower left panel). The results of the global fit
vs.\ the observed $\beta$ color is presented in the lower right panel.
Symbols are as in Fig.~1, except that two curves 
are presented for MARCS models: the shallower curve
is the one computed by us while the other one is from O09. Neither curve
is able to explain the low $\beta$ colors observed for [Fe/H] $< -0.2$.
}
\label{betasyn}
\end{figure}

Although there is a clear trend of $\beta$ vs.\ V magnitude
(Fig. 2), it is not related to reddening because the
$\beta$ index is not affected by interstellar absorption.
The bright stars (V $<$ 6) that are causing the 
trend are within $\sim$17 pc, hence they are not reddened.
The bright (V $<$ 6) solar twins and analogs falling
systematically above the derived $\beta_\odot$, are
stars hotter than the Sun, or more metal-rich than
the Sun (or both), and therefore with 
a $\beta$ color  systematically higher than solar
because $\beta$ increases with both increasing metallicity and 
\tsin. We note that the trend seen in nearby stars for
$\beta$ is not present in $(b$--$y)$ because this color 
has opposite trends with \teff and [Fe/H].
Thus, by performing a deeper solar-twin survey than 
previous works, we avoided systematic biases (such as selecting
mainly hotter and more metal-rich stars) that 
might have been present if only brighter stars were studied.

\subsection{The $m_1$ solar color}
It is well known that the $m_1$ color correlates very well with metallicity
(see references in the introduction), 
and this is clearly shown in Table 4, 
where according to the global fit
the dependence on [Fe/H] is significant at the 21-$\sigma$ level.
The second most important parameter is \teff (10 $\sigma$),
but log $g$ also produces an important dependence
(slope significant at the 5-$\sigma$ level).

The tight correlation between $m_1$ and [Fe/H] is shown in Fig.~3. 
The star-to-star scatter from the global fit is
only 0.006 magnitudes, which is compatible with the observational
uncertainties. The $m_1$ solar color recommended for the Sun is

\begin{equation}
m_{1, \odot} = 0.2122 \; (\pm 0.0018).
\end{equation}

\begin{figure}
\resizebox{\hsize}{!}{\includegraphics{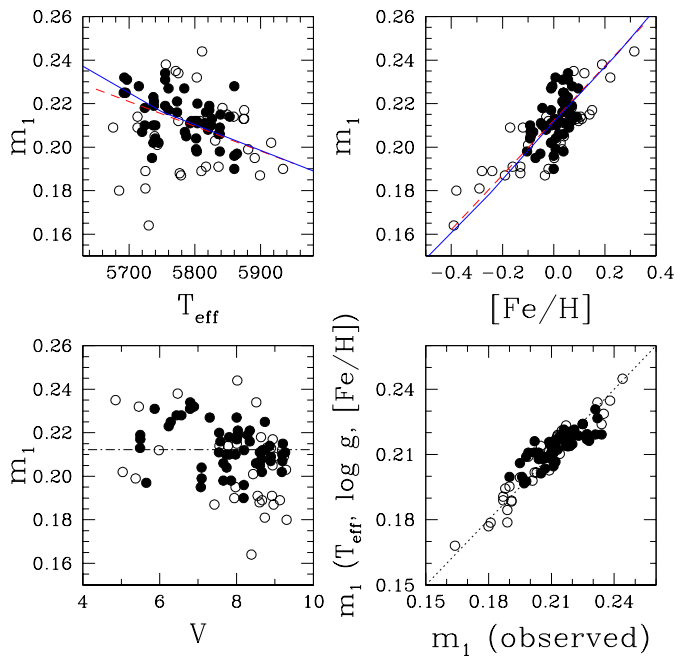}}
\caption{$m_1$ vs.\ \teff (upper left panel), [Fe/H] (upper right panel),
and V magnitude (lower left panel). The results of the global fit
vs.\ the observed $m_1$ color is presented in the lower right panel. 
Symbols are as in Fig.~1.}
\label{m1syn}
\end{figure}

\subsection{The $c_1$ solar color}

In Fig.~4, we show the relation between $c_1$ and log~$g$, 
which is significant at the 4-$\sigma$ level.
As already noticed in the literature \citep[e.g.][hereafter O09]{twa02,one09},
the $c_1$ index in late G dwarfs has a sensitivity to metallicity.
Our global fit (Table 4) confirms the dependence on
[Fe/H], which is actually more significant (8-$\sigma$)
than the dependence on log~$g$ (Fig.~4).

The predicted $c_1$ solar color from the global fit is

\begin{equation}
c_{1, \odot} = 0.3319 \; (\pm 0.0054).
\end{equation}

\noindent The star-to-star scatter in the global fit is
0.0165, which is considerably larger than the 
observational errors (0.009 - 0.013) for the $c_1$ index.
Following the suggestions of the referee, we
explored whether this index is particularly sensitive
to either anomalies in the chemical composition or the
microturbulence velocities. 

Previous works show the effects of C and N upon 
the Str\"omgren 4-color $uvby$ filter-measurements, 
via the NH and CN bands. For example, it seems 
that the NH band at 3360A affects the $u$ measurements 
while CN affects the $v$ measurements. Thus, the $c_1$ color
that depends on both $u$ and $v$ should be affected
by abundance anomalies
\citep{bon69,bon74,zac98,gru00,gru02,sch04,yon08}.
In this context, it would be important to assess
whether the small abundance anomalies in the solar chemical composition
(M09; R09),
in particular the difference between the highly volatile elements (C, N, O)
and Fe, may affect the $uvby$ colors.
As discussed by \cite{str82}, variations in the He abundance
may also affect the $c_1$ index. Although the Hyades $c_1$
anomaly for stars with $(b-y)$ close to solar was found initially
to be $\Delta c_1 \sim$0.03-0.04 (Hyades - field stars, or Hyades - Coma),
it seems that the anomaly may only amount to $\Delta c_1$=0.024-0.025 
after instrumental effects are corrected \citep{jon95}.
Another important parameter affecting the colors may be microturbulence
\citep{con66,nis81}.

To test the above effects, synthetic spectra were computed
for solar twins with variations in $\Delta$[C,N,O/Fe]=$-$0.05 dex
(M09), $\Delta$v$_{\rm micro}$ = +0.1 km s$^{-1}$ (most
solar twins and close solar analogs have v$_{\rm micro}$ within
$\pm$0.1 km s$^{-1}$ of the solar value), and an increase of 10\% in
the He abundance (by number). Fluxes were computed with the code SYNTHE
\citep{kur81,sbo04,kur05} using ATLAS12 model atmospheres
\citep{kur96,kur05,cas05} computed for the different aforementioned assumptions.  
The atomic line list adopted in the spectral
computations is based on the compilations by \cite{coe05} and
\cite{cash04}, and the molecules C$_2$, CH, CN, CO, H$_2$, MgH, NH, OH,
SiH, and SiO from \cite{kur93} were included.  The change in He
does not significantly affect the $c_1$ index ($<$0.001 magnitudes), but
the change in microturbulence increases $c_1$ by +0.0035 mag., while
the change in C,N,O increases $c_1$ by +0.003 mag.\footnote{The
changes in $(b-y)$ and $m_1$ due to changes in He, CNO, and microturbulence
are even smaller than for $c_1$.} Thus, small
changes in chemical composition and microturbulence may help to
explain the extra scatter seen in the global fit. 
Casagrande et al.\ (in preparation) demonstrate that the
Str\"omgren system is not only sensitive to [Fe/H], but that it
is possible to obtain information about the [$\alpha$/Fe] ratio using
$uvby$ photometry.  The $\Delta c_1$ anomaly in the
Hyades could be due to the effect of metallicity 
on $c_1$.  According to \cite{fri92}, Hyades has
an iron abundance 0.18 dex higher than Coma, which 
according to Table 4 corresponds to $\Delta c_1 =$
+0.022 mag., which is very close to the Hyades anomaly
relative to Coma (0.024-0.025 mag.) \citep{jon95}.

\begin{figure}
\resizebox{\hsize}{!}{\includegraphics{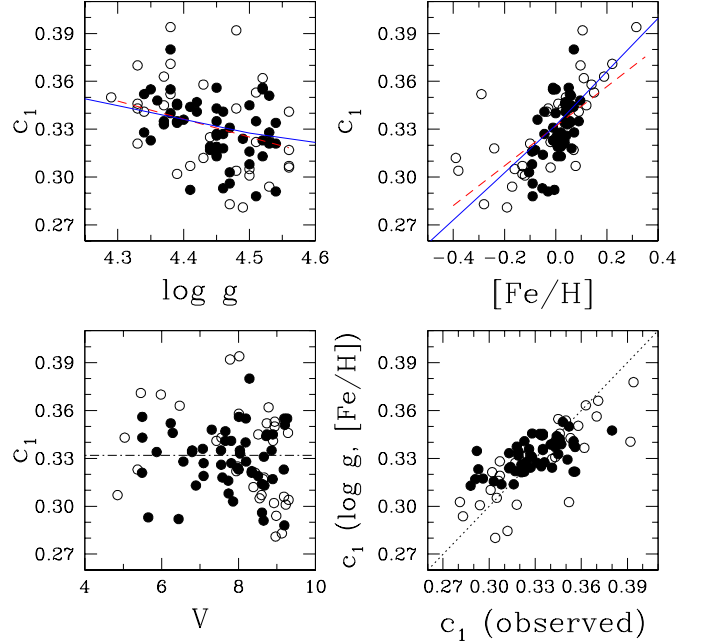}}
\caption{$c_1$ vs.\ log $g$ (upper left panel), [Fe/H] (upper right panel)
and V magnitude (lower left panel). The results of the global fit
vs.\ the observed $c_1$ color is presented in the lower right panel. 
Symbols are as in Fig.~1.}
\label{c1syn}
\end{figure}

\section{Comparison with previous empirical determinations}

As can be seen in Table 5 (lower part), our inferred $(b-y), m_1, c_1$, and
$\beta$ solar colors agree well (within 1-$\sigma$) with most earlier
empirical results.  However, previous determinations of solar
colors have much lower control and homogeneity in \tsin, log $g$, and [Fe/H],
thus in some cases the agreement (within the error bars) could be due to
fortuitous compensations of \tsin, log $g$, and [Fe/H] effects on colors, or
due to large errors in other studies.

Several previous works \citep{sax85,gra92,edv93,cas10} 
give $(b-y)_{\odot}$ values that agree with ours within 0.005 magnitudes,
but other determinations are either too blue (\citealp{cay96,cle04}; RM05b; \citealp{hol06})
or too red \citep{geh81}.
A previous empirical result for $\beta_{\odot}$
\citep[][$\beta_{\odot} = 2.591\pm0.005$]{sax85}
is in excellent agreement with our value (2.5915$\pm$0.0024).
Previous determinations of $m_{1, \odot}$ and $c_{1, \odot}$
\citep{cle04,hol06} also agree with ours within the errors, 
but note that the errors given by \cite{hol06} are very large
(0.03 and 0.07 for $m_1$ and $c_1$, respectively), so their
solar colors cannot be used to perform stringent tests of model atmospheres
and the temperature and metallicity scales.

Although our inferred solar colors agree with previous determinations based
on solar-type stars, our values based on solar twins should be preferred 
because of their more accurate and precise stellar parameters, 
resulting in correspondingly accurate and precise inferred solar colors.
We use our ``solar'' colors below to test absolutely calibrated
solar spectra, model atmospheres and the \teff and metallicity $uvby$-$\beta$
scales.

\section{Comparison with measurements on reflected (asteroid) solar spectra}

\cite{ols76} measured the $\beta_{\odot}$ color of reflected solar
light from asteroids, which seems justifiable because the
$\beta$ index should be largely independent of the 
wavelength dependence of the albedo. His measurement
($\beta_{\odot} = 2.5955\pm0.0024$) is in good agreement (within 1-$\sigma$)
with our inferred value ($\beta_{\odot} =  2.5915\pm0.0024$).
\cite{ols76} did not measure the solar $(b$--$y)$, but 
indirectly estimated $(b-y)_{\odot}$ from a color
transformation using $\beta_{\odot}$. His $(b-y)_{\odot} = 0.390\pm0.004$
disagrees with ours (0.4105$\pm$0.0015), but this is probably due to the
errors introduced by his adopted transformation from $\beta$ to $(b$--$y)$.
Using the $(b-y)_0$--$\beta$ intrinsic color calibration 
of \cite{kar10}, the $\beta_{\odot} = 2.5955$ from
\cite{ols76} indeed implies that $(b$--$y)_{\odot} = 0.4062$, in good
agreement with our $(b$--$y)_{\odot}$. Using the same transformation,
our $\beta_{\odot} = 2.5915$ gives $(b$--$y)_{\odot} = 0.4089$.

Unfortunately, direct measurements of the $(b$--$y)$, $m_1$ and $c_1$ color
indices for reflected solar spectra are not very useful because of the
color of the asteroid albedos\footnote{Although the colors of the Sun measured
using asteroids may be affected, our spectroscopic analysis based on 
high-resolution spectra should not because we measure the flux
relative to the adjacent continuum for narrow spectral lines}.
A comparison with solar colors inferred
from absolute spectrophotometry of the Sun is presented below.

\section{Comparison with synthetic colors}

Our accurate and precise colors inferred for the Sun can be used to test the
performance of theoretical solar flux models and the quality of absolute
solar flux measurements. However, additional ingredients enter the computation
of synthetic colors, namely the adopted set of filters and the flux/magnitude
zero-points adopted. We discuss them in the following.

Since intermediate-band Str\"omgren filters are centered on specific spectral
features, a correct characterization of the total throughout becomes crucial
for generating synthetic colors (\citealp[e.g.][]{les86}; O09).
To test the influence of the filter transmission curves on our
results, we computed $uvby$ indices using two different sets of pass-bands,
the original ones of \cite{cra70} and a set that should be more representative
of the SPM observations \citep{bes05}.

The $\beta$ index is defined as the ratio of the flux measured through narrow
(half-width of about $30$\,\AA) and wide (about $150$\,\AA) profiles both
centered on the H$\beta$ line. In this case, the (212,214) filter transmission
curves (Crawford \& Mander, 1966), the photomultiplier sensitivity, the
atmospheric transmission, and the reflectivity of aluminum given in
\cite{cas06} were used to generate the total $T_\beta$ throughout,
according to the prescriptions of the beta.forcd program at the Kurucz
website\footnote{http://kurucz.harvard.edu/programs/COLORS}. Indices
calculated using $T_\beta$ define the natural system $\beta'$, which should be
transformed using a set of equations to agree to the standard system
$\beta$ defined by the observations of \cite{cra66}. For the filter set
(212,214), the transformation equation is $\beta=0.248+1.368 \, \beta'$
\citep{cra66}.

In principle, to mimic the $uvby$-$\beta$ observations presented here,
synthetic photometry should reproduce the SPM instrumental system, and the same
transformation equations should then be applied to generate the standard system.
In practice, this can hardly be done, since the SPM instrument is a six-channel
spectrophotometer, and we follow the approach normally adopted in the
literature, i.e.\ of reproducing the standard system directly, by fixing the 
zeropoints using Vega. In Sect. 2.4, we have shown the excellent agreement between our
observations and other photometric measurements, meaning that the transformation
from the instrumental to the standard system is indeed accurate; we therefore
expect our approach to return meaningful synthetic colors.

The spectral energy distribution adopted for Vega and its observed
indices also affect the outcome of synthetic photometry in the process of
establishing the zeropoints of a photometric system
\citep[e.g.][]{casagrande06}. We used a spectrum obtained with the
STIS spectrograph onboard the HST \citep{bohlin07} of resolution
$\mathcal{R} = 500$, which enables the highest accuracy ($\sim 1$\%) measurements
achievable to date, and adopted the following averaged values for Vega:  $(b-y)=0.003$,
$m_1=0.157$, $c_1=1.088$ and $\beta=2.904$ from
\cite{hau98}. If we had adopted the colors of Vega given in
\cite{cra72}, the differences would have amounted to $0.001$ magnitude at most.

Despite the complications posed by the pole-on and rapidly rotating nature of
Vega, the effects on the blue part of the spectrum are expected to be small or
negligible \citep[e.g.][and references therein]{casagrande06,bohlin07}.
It is more relevant that we use the observed HST spectrophotometry of
Vega; additional comments on this issue are made in the following subsections.

\subsection{Colors from Kurucz and MARCS models}\label{sec:synthe}

Although synthetic $uvby$-$\beta$ colors computed using earlier MARCS and
Kurucz models are found in the literature (see Table 5), we feel it appropriate
to make our comparisons using the most recent releases available
\citep{cas04,gus08}. In addition,
as we have already mentioned, different ingredients enter the computation of
synthetic colors, and we differ from most of the previous works in that we use
an observed spectrum of Vega to define the zeropoints. In principle, this should
be the best approach for replicating observations and for a correct comparison
of solar flux models and absolute measurements with the colors determined
from our solar twins. This choice mimics the observational approach,
and the successes or failures of synthetic colors depend mostly on the quality
of the solar input spectra.  In practice, the situation is less clear, as we
discuss further below and in Sect. \ref{sec:absflx}.

When computing synthetic colors from theoretical models, the use of a
model atmosphere to describe also Vega may have the advantage of (partly)
compensating for model inaccuracies by including these in the zero-points.

We performed this exercise by taking from the Kurucz website the latest ATLAS9
model fluxes for Vega and the Sun; because of the internal consistency of
this approach, it should also determine which set of filters should be
used.  For the colors of the Sun
we obtain $(b-y)=0.413$, $m_1=0.236$ and $c_1=0.297$ using the filters of \cite{cra70},
and $(b-y)=0.406$, $m_1=0.214$, $c_1=0.303$ for the \cite{bes05} filters.
While $(b$--$y)_\odot$ is reproduced with the two set of filters, the latter set of filters
provides results that are more comparable
to our measured $m_{1, \odot}$ and $c_{1, \odot}$, and therefore in
the following we consider only the \cite{bes05} passbands. The effect of
instead using those of \cite{cra70} can be easily estimated from the above
differences.

The resolution of the spectra used to generate synthetic colors may also in
principle affect the results.  We tested ATLAS9 model fluxes for Vega and the
Sun at various resolutions ranging from $\mathcal{R}=500000$ to $200$ and
verified that for various combinations of these, differences in $(b$--$y)$,
$m_1$ and $c_1$ always lie below $0.001$ magnitude.\footnote{The \cite{cas04}
grid of fluxes has a resolution varying from $150$ to $250$ in the wavelength
region of interest, and for this reason we took instead model fluxes at
higher resolution from the Kurucz website} 

However, for $\mathcal{R} \lesssim 2000$ the synthetic $\beta$ index
scales differently for Vega and the Sun, i.e.\ even if synthetic spectra of
the same resolution are used for the two stars, the value obtained for $\beta$
depends on $\mathcal{R}$ to an extent that may vary from a few
milli-magnitudes up to several hundredths of a magnitude.  For example,
using high resolution ATLAS9 spectra for both Vega and the Sun, we obtain
$\beta=2.587$, which is in good agreement with our solar value. Using instead
synthetic spectra of Vega and the Sun at $\mathcal{R}=500$ sampled at the
same wavelength points, we obtain $\beta=2.601$ from which we estimate a 
(model-dependent) correction of $0.014$ magnitude when working at this low
$\mathcal{R}$.

\begin{figure*}
\resizebox{\hsize}{!}{\includegraphics{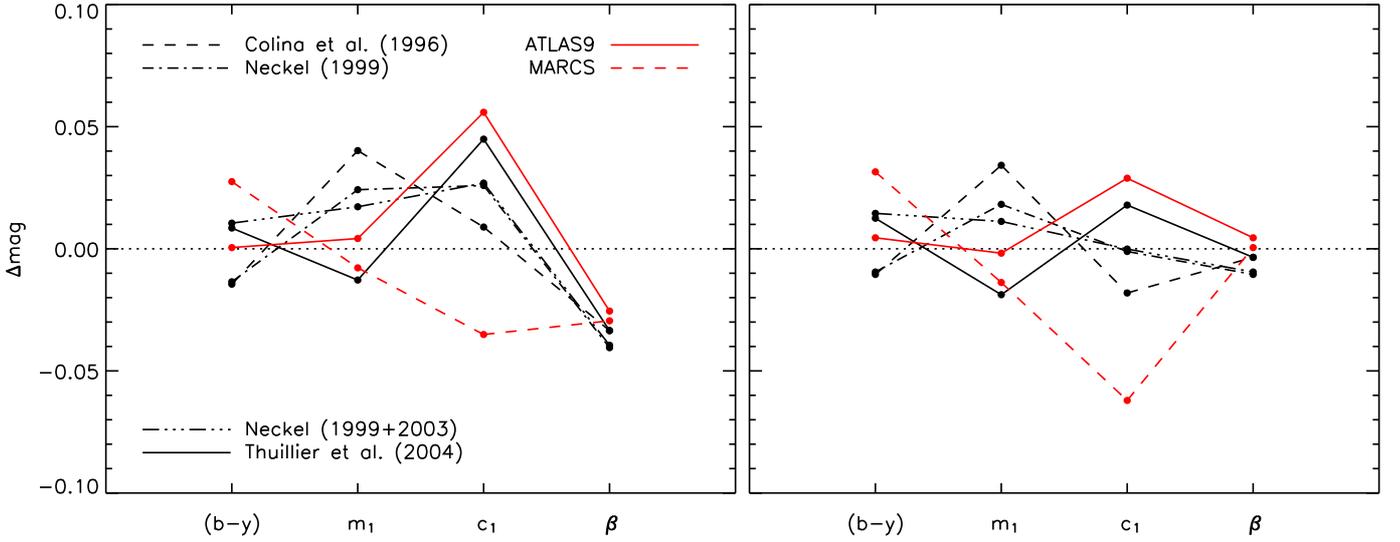}}
\caption{Comparison between solar and synthetic colors obtained from different 
spectra. Left panel: using Vega STIS observations to set the zero-points. 
Right panel: using ATLAS9 Vega model for the same purpose.  $\Delta$mag =
our derived solar colors minus the synthetic ones.}
\label{delta_col}
\end{figure*}

Because of this limitation on the STIS resolution of the Balmer line, the
$\beta$ indices computed for Table 5 have been obtained as follows:  the solar
spectra were downgraded to $\mathcal{R}=500$, sampled at the same wavelength
points as the STIS spectrum, and the aforementioned correction was then applied.
Using instead a high resolution ATLAS9 spectrum of Vega to define the zero-points
does not require a downgrade to the resolution of the solar spectra, and the
$\beta$ indices are changed by $-0.030$ magnitude (i.e.\ are bluer).
We verified that for the other Str\"omgren indices the STIS resolution is
high enough. The effect of using an
ATLAS9 model flux of Vega to set the zero-points amounts to
$-0.004$, $0.006$, and $0.027$ magnitudes for the $(b$--$y)$, $m_1$, and $c_1$
indices, respectively. In Figure \ref{delta_col}, synthetic colors obtained
with these two choices for Vega, i.e.\ STIS vs.\ ATLAS9, are compared.

If the STIS spectrum of Vega is used, the Kurucz solar flux returns 
a value of $(b$--$y)$ in
excellent agreement with our observed value, whereas a considerably bluer
color is obtained with MARCS. The greater accuracy of the Kurucz with
respect to the MARCS solar flux in $b$ and $y$ bands can be noticed also from
the spectrophotometric comparison presented in \cite{edv08}.
For $m_1$, the Kurucz model is still in better agreement than the MARCS, one 
being slightly bluer and the other redder with respect to our $m_{1, \odot}$. The
same bluer and redder performance is also obtained for $c_1$, although
both models provide a rather poor match.  The $\beta$ index is similar for
both models and considerably redder than our $\beta_{\odot}$.

In general, the ATLAS9 solar flux model performs better than the MARCS model, but 
they both have problems in reproducing $c_{1, \odot}$ (in opposite directions) and 
$\beta_\odot$ (both systematically redder). Changing to the ATLAS9 flux
of Vega to set the zero-points has a negligible impact on $(b$--$y)$ and $m_1$, 
but brings the theoretical $\beta$ index in almost perfect agreement in both 
cases. For the $c_1$ index, only the ATLAS9 result is helped by this 
choice, and this could partly stem from compensating errors in the ATLAS9 
models of Vega and the Sun.

In Table 5, we also show the color indices computed by O09
using MARCS models (also for the flux of Vega) and the passbands of
\cite{cra70}.  Their $(b$--$y)$ agrees with our synthetic one, but their $m_1$
and $c_1$ colors are somewhat redder and bluer, respectively, in a way
that is consistent with
the different passbands they adopted. The $c_1$ solar color computed by
O09 is in excellent agreement (within 1-$\sigma$) with our $c_{1, \odot}$.

\subsection{Colors from solar spectra}\label{sec:absflx}

We computed $(b$--$y)$, $m_1$, $c_1$, and $\beta$ indices using the absolute 
measurements of solar spectra by C96, Neckel (1999, hereafter N99), and \cite{thui04}.
The spectrum of \cite{rie08} was not employed because of its very low resolution 
($R\sim100$). The STIS Vega observations and the ATLAS9 Vega model
were again used to define the zero points.

The C96 composite spectrum represents both satellite \citep{woo96} and ground
\citep{nec84} observations below and above $4100$\,\AA.  While its $(b$--$y)$ is
redder, $m_1$ is considerably bluer, and $c_1$ is in good agreement
with our solar values. Not unexpectedly, the same conclusions about $(b$--$y)$ and
$m_1$ hold for the N99 atlas, which is an update of the \cite{nec84} measurements
included in C96.  However, Neckel (2003, hereafter N03) noticed a possible
systematic error in those absolute measurements and provides a simple analytical
formula for correction, after which $(b$--$y)$ and $m_1$ are in closer agreement
with our solar values.  \cite{thui04} published two composite solar reference
spectra assembled using space measurements during distinct solar activity levels;
differences concern only the $m_1$ and $c_1$ indices in a negligible manner (a
few millimagnitudes only), and therefore in Table 5 the averaged values of the
two are given.  \cite{thui04} provide closer agreement than previous spectra
in $(b$--$y)$ and $m_1$, but their $c_1$ is considerably bluer.

Thus, while on average $(b$--$y)$ and $m_1$ are in agreement with our solar
values, there is the tendency for $c_1$ to be considerably bluer. The $\beta$
indices were computed following the same prescriptions as in the
previous section and are systematically redder. Switching to the model flux of
Vega solves most of the discrepancies for these two indices, with the
synthetic values being now distributed at the red and blue sides of our solar
colors (see the right panel of Fig.~5). It is not obvious why a model flux for
Vega also brings absolute solar flux measurements into closer agreement with
our empirical solar colors.  While the measured spectra of Vega should be
superior to any model of it, we can speculate about possible reasons for this
not being true over the full wavelength range. The difference STIS vs.\ ATLAS9
amounts to few millimagnitudes in $(b$--$y)$ and $m_1$, but is considerable in
$c_1$ and $\beta$, and those differences could reflect measurement problems
in the $u$ band and the issue already mentioned about the spectral resolution
around the Balmer line.

In the wavelength range of interest to us, the accuracy of the solar absolute
fluxes is on the order of $5-3$\% from data acquired in space (C96; \citealp{thui04}) and 
probably lower for ground-based data measurements. The culprit could thus be inaccuracy in
solar flux measurements rather than in Vega!  For the sake of computing
indices, we are not interested in the absolute flux scale, but rather in the
relative accuracy of measuring the shape of the solar spectral energy
distribution.  This accuracy is rather difficult to assess, but the above
differences provide an idea of the complexity associated with this kind of
measurement, and the corresponding rather large uncertainties. In addition,
the choice for Vega and the set of filters used also affect the analysis,
introducing systematic errors.

In this section, we have presented various issues concerning synthetic
solar colors, which are often taken for granted when comparing observed
vs.\ synthetic colors. While the comparison with a specific solar-flux
measurement has limited significance, with appropriate choices for Vega,
it is comforting that there are no large systematic trends between our
solar colors and these observations taken together. Thus, our solar colors
can help considerably in establishing the necessary reliability of
synthetic colors.

\subsection{Testing the effects of changes in \tsin, log $g$, and [Fe/H] on
synthetic MARCS colors}

As discussed in the previous sections, our $uvby$-$\beta$ solar colors
are useful for testing absolute solar spectra and the performance of model
atmospheres, as well as checking the subtleties in the transformation from
fluxes to synthetic colors. Since our solar twins and analogs
span a range in \tsin, log $g$, and [Fe/H], we can also use them to
test how well model atmospheres predict the relative change of colors due
to variations in atmospheric parameters.

For this test, we used relative fluxes predicted by the MARCS models,
normalizing the results to our ``observed'' solar colors.
As shown in Figs. 1--4, where the relative variation in the
MARCS $(b$--$y)$, $\beta$, $m_1$, and $c_1$ colors are shown by 
solid lines, 
there is a satisfactory overall {\em relative} agreement between MARCS models 
and both the solar twins and solar analogs.  The only clear discrepancy is
in the predicted metallicity variation in $\beta$ (Fig.~2), which is much
shallower than observed. We also checked whether the MARCS $\beta$
colors computed by O09 solve this problem (Fig.~2), but although they do help
to slightly alleviate the discrepancy (the \"Onehag et al.\ $\beta$ colors
are slightly steeper than ours), they also do not match the low $\beta$ indices
observed below [Fe/H] $\sim -0.2$.
It would be important to observe more
metal-poor solar analogs to confirm whether we have detected a potential problem in
the line formation of H$\beta$.

The strong variation in $c_1$ with metallicity for G-type stars
(\citealp[Fig.~4; see also]{twa02} O09) is reproduced well by the MARCS models
(as already demonstrated by O09 for somewhat cooler dwarfs), so it can
potentially be used to estimate [Fe/H] in 
solar metallicity
and metal-rich solar analogs.

Our tests provide confidence in the relative variation in stellar fluxes
with atmospheric parameters as predicted by the MARCS models,
at least in the atmospheric parameter space studied here.

\section{The zero-point of the \teff scale}

Our solar $(b$--$y)$ color can also be used to estimate how well
different \teff scales reproduce the solar value of \teff (5777 K).
Different color-\teff relations in the literature have been employed
to obtain the \teff for our $(b-y)_{\odot}$, and these values
have then been subtracted from the solar effective temperature. These
differences, shown in Table 6, are the zero-point corrections needed
to place different \teff calibrations on the same \teff scale as the Sun.
As can be seen, the \teff scale by \cite{cas10} seems the most
accurate, with a negligible offset of only 
10($\pm$10) K with respect to the Sun,
which is much smaller than the offset (109 K)
of our earlier \teff calibration (RM05b).

The widely used \teff $(b$--$y)$ scale of \cite{alo96} 
is too cool by 137 K.  Thus, the elemental chemical abundances
determined using this scale \citep[e.g.][]{isr98,che00,red03,all04,jon05}
may be systematically incorrect by as much as 0.14 dex (e.g. for V from VI lines and for N from
NH lines), albeit the impact on abundance ratios [X/Fe] should be smaller
(and in a few cases mostly cancel) for lines of low to moderate excitation
potential because of the compensating impact of the iron abundance obtained
from FeI lines; however, for abundances obtained from lines of high excitation 
potential (e.g. CI, NI, OI, PI, SI) the impact could be as high as 0.2 dex.
Thus, inaccurate temperature scales may be one of the reasons why the small
peculiarities in the solar chemical composition (M09; R09) have
not been discovered in the past \citep{ram10}.

Zero-point errors in \teff also affect stellar ages inferred from isochrones.
For example, for an offset of 130 K and using Y$^2$ isochrones \citep{dem04},
a systematic offset as high as 1.5-2 Gyr may result for a solar analog.


\section{The zero-point of the metallicity scale}

\cite{arn10} made a critical evaluation of different
metallicity calibrations for dwarf stars in the $uvby$ system, finding zero-point
offsets in many of these metallicity scales, ranging from $-0.17$ dex \citep{nor04}
to +0.33 (O09), although the latter is a purely theoretical calibration
intended to test model atmospheres and not for application to real stars.
After extensive testing for potential problems in different metallicity
calibrations, such as trends in the $\Delta$[Fe/H] residuals
(spectroscopic-photometric) with [Fe/H], $(b-y), c_0$, and $m_0$, \cite{arn10} find
that the calibrations by RM05a has the best overall performance, albeit
with a small (0.04 dex) zero-point offset.

The zero-point offset cannot be overstated, especially when comparing the Sun
to the stars \citep{gus_sun08}, because offsets in the zero-point of the metallicity
scale could make the Sun appear abnormal (see Haywood, 2008; Gustafsson et al.,
2010; and references therein for a discussion about apparent anomalies in the solar
metallicity).

Our solar colors can be used to check the zero-point of different metallicity
calibrations, by computing [Fe/H] for our inferred solar colors and subtracting
these metallicities from the solar metallicity ([Fe/H] = 0). Those differences,
which are the zero-point corrections needed to place different metallicity
calibrations on the same metallicity scale as the Sun,
are shown in Table 7, along with the offsets found by
\cite{arn10} for a broad range of colors and metallicities.

Most metallicity calibrations do require a correction to place them
on the same metallicity scale as the Sun.  In particular,
the RM05a metallicity calibration
recommended by \cite{arn10} to derive [Fe/H] from Str\"omgren
photometry, needs a zero-point correction of 
+0.04 dex. This offset is identical to
the offset found by
\cite{arn10} using a sample spanning a much broader range in \tsin.
We are currently revising the metallicity calibration employed
in the GCS survey \citep{nor04,hol07} and plan to assess the
apparent anomalously high metallicity of the Sun with
a new accurate metallicity scale (Casagrande et al., in preparation).

\section{Conclusions}

New $uvby$-$\beta$ photometry has been presented for solar-twin candidates
observed at the SPM observatory.  Comparisons with existing Str\"omgren
photometry shows that our data are in excellent agreement (to better than
0.01 mag) with previous observations.

Using accurate spectroscopically derived stellar parameters, the
$uvby$-$\beta$ photometry for the solar twins, and also for solar analogs
covering a wider range in metallicities, the solar colors
$(b-y)_{\odot}$ = 0.4105$\pm$0.0015, $m_{1, \odot}$ =  0.2122$\pm$0.0018,
$c_{1, \odot}$ = 0.3319$\pm$0.0054,
and $\beta_{\odot}$= 2.5915$\pm$0.0024
have been inferred.

As discussed in the manuscript, our solar-twin data have provided stringent
constraints on absolute fluxes, the performance of model
atmospheres, and the zero-points of temperature and metallicity scales.
In particular, we show that the widely used \cite{alo96} $(b$--$y)$
calibration is too cool by $\sim$140 K, while our new \teff calibration
\citep{cas10} has a negligible zero-point offset 
(10$\pm$10 K).
Regarding model atmospheres, the Kurucz ATLAS9 solar model provides 
a closer fit of our solar colors than the MARCS 2008 solar model.  The
relative variation in colors with stellar parameters seem to be
reproduced well by MARCS models, except for the metallicity variation
in the $\beta$ index, which is flatter than observed in solar
twins and solar analogs, thus suggesting that there are existing
limitations on the modeling of Balmer lines.

We are pursuing photometric observations of our solar-twin sample in other
photometric systems, to perform similar evaluations of absolute
fluxes, model atmospheres, and fundamental calibrations in astrophysics.

\begin{acknowledgements}
We thank A. \"Onehag and B. Gustafsson for computing the solar color indices
(in the O09 system) given in Table 5, and also by comments on our
results.  W.J.S. is very grateful to the DGAPA--PAPIIT (UNAM) 
(projects Nos. IN101495, and IN111500) and to CONACyT (M\'exico) 
(projects Nos. 1219--E9203, 27884E, and 49434F) for funding which permitted
travel and also the maintenance and upgrading of the $uvby$--$\beta$ photometer.
J.M. would like to acknowledge support from Funda\c{c}\~ao para a
 Ci\^encia e a Tecnologia (FCT, Portugal) in the form of a grant
 (PTDC/CTE-AST/098528/2008) and a Ci\^encia 2007 fellowship.
P.C. is grateful to F. Castelli, P. Bonifacio and L. Sbordone 
for their long term help with the SYNTHE and ATLAS12 codes.
This publication has made use of the SIMBAD database, operated at CDS,
Strasbourg, France.
\end{acknowledgements}

\clearpage

\onltab{1}{
\scriptsize
\begin{table}
\caption{Photometry (V Johnson and $uvby$-$\beta$) of 73 solar-twin candidates from San Pedro M\'artir Observatory}
\label{tab1}
\centering 
\renewcommand{\footnoterule}{}  
\begin{tabular}{rrrrrrrrrrrrrrrrrrrrrrrrrrrr} 
\hline
\hline 
{HIP number}  & HD/BD number & V & $(b$--$y)$ & $m_1$ & $c_1$ & N$_V$ & N$_{uvby}$ & $\beta$ & N$_\beta$ \\
\hline
348    & 225194   & 8.600 & 0.407 & 0.188 & 0.307 & 4 & 4 &   2.584 & 4 \\
996    & 804      & 8.184 & 0.403 & 0.190 & 0.340 & 4 & 4 &   2.598 & 4 \\
1411   & 1327     & 9.088 & 0.441 & 0.211 & 0.342 & 4 & 4 &   2.579 & 4 \\
2894   & +48 0182 & 8.653 & 0.415 & 0.208 & 0.291 & 4 & 4 &   2.587 & 4 \\
4909   & 6204     & 8.508 & 0.403 & 0.206 & 0.319 & 4 & 4 &   2.597 & 4 \\
5134   & 6470     & 8.965 & 0.402 & 0.187 & 0.281 & 4 & 4 &   2.581 & 4 \\
6407   &  8291    & 8.612 & 0.411 & 0.205 & 0.296 & 4 & 4 &   2.584 & 4 \\
7245   &  9446    & 8.377 & 0.423 & 0.218 & 0.328 & 4 & 4 &   2.596 & 4 \\
8507   & 11195    & 8.890 & 0.413 & 0.207 & 0.317 & 4 & 4 &   2.595 & 4 \\
8841   &$-$13 0347& 9.232 & 0.423 & 0.209 & 0.301 & 4 & 4 &   2.593 & 4 \\
9349   &  12264   & 7.975 & 0.407 & 0.210 & 0.324 & 4 & 4 &   2.594 & 4 \\
10710  & +18 0289 & 8.924 & 0.400 & 0.191 & 0.302 & 5 & 5 &   2.594 & 5 \\
11728  & 15632    & 8.041 & 0.414 & 0.219 & 0.333 & 5 & 5 &   2.590 & 5 \\
11915  & 16008    & 8.608 & 0.406 & 0.211 & 0.314 & 4 & 4 &   2.594 & 4 \\
14614  & 19518    & 7.858 & 0.400 & 0.198 & 0.303 & 5 & 5 &   2.591 & 5 \\
18261  & 24552    & 7.976 & 0.390 & 0.195 & 0.325 & 5 & 5 &   2.604 & 5 \\
25670  & 36152    & 8.285 & 0.413 & 0.219 & 0.380 & 6 & 6 &   2.599 & 6 \\
28336  & 40620    & 8.982 & 0.411 & 0.209 & 0.294 & 6 & 6 &   2.578 & 6 \\
35265  & 56124    & 6.936 & 0.395 & 0.207 & 0.349 & 7 & 7 &   2.594 & 7 \\
36512  & 59711    & 7.732 & 0.403 & 0.210 & 0.316 & 5 & 5 &   2.588 & 5 \\
38072  & 63487    & 9.206 & 0.406 & 0.215 & 0.351 & 6 & 6 &   2.601 & 6 \\
41317  & 71334    & 7.808 & 0.409 & 0.218 & 0.341 & 5 & 5 &   2.588 & 5 \\
41832  & 71779    & 8.125 & 0.399 & 0.199 & 0.322 & 7 & 7 &   2.580 & 7 \\
42872  & 74645    & 9.276 & 0.402 & 0.198 & 0.363 & 4 & 4 &   2.594 & 4 \\
44324  & 77006    & 7.942 & 0.390 & 0.190 & 0.322 & 6 & 6 &   2.587 & 6 \\
44935  & 78534    & 8.722 & 0.413 & 0.212 & 0.344 & 5 & 5 &   2.590 & 5 \\
44997  & 78660    & 8.346 & 0.412 & 0.221 & 0.322 & 5 & 5 &   2.588 & 5 \\
47990  & 84705    & 8.704 & 0.406 & 0.218 & 0.327 & 5 & 5 &   2.591 & 5 \\
49572  & +30 1962 & 9.288 & 0.406 & 0.203 & 0.346 & 5 & 5 &   2.591 & 5 \\
49756  & 88072    & 7.549 & 0.403 & 0.220 & 0.326 & 5 & 5 &   2.593 & 5 \\
50826  & +17 2213 & 9.129 & 0.403 & 0.189 & 0.283 & 5 & 5 &   2.581 & 5 \\
51337  & 90733    & 8.905 & 0.399 & 0.203 & 0.354 & 5 & 5 &   2.595 & 5 \\
52040  & 91909    & 9.194 & 0.406 & 0.207 & 0.288 & 5 & 5 &   2.573 & 5 \\
52137  & 92074    & 8.638 & 0.415 & 0.218 & 0.317 & 5 & 5 &   2.584 & 5 \\
55409  & 98649    & 8.010 & 0.407 & 0.217 & 0.356 & 5 & 5 &   2.585 & 5 \\
55459  & 98618    & 7.658 & 0.406 & 0.206 & 0.341 & 5 & 5 &   2.607 & 5 \\
58303  & 103828   & 8.443 & 0.411 & 0.210 & 0.314 & 5 & 5 &   2.642 & 5 \\
59357  & 105779   & 8.665 & 0.389 & 0.189 & 0.318 & 5 & 5 &   2.581 & 5 \\
60314  & 107633   & 8.783 & 0.409 & 0.213 & 0.362 & 4 & 5 &   2.592 & 5 \\
60653  & 108204   & 8.737 & 0.401 & 0.181 & 0.352 & 5 & 5 &   2.579 & 5 \\
63048  & 112257   & 7.808 & 0.423 & 0.220 & 0.351 & 5 & 5 &   2.582 & 5 \\
64150  & 114174   & 6.795 & 0.412 & 0.234 & 0.335 & 5 & 5 &   2.584 & 5 \\
64497  & 114826   & 8.943 & 0.411 & 0.215 & 0.351 & 5 & 5 &   2.598 & 5 \\
64713  & 115169   & 9.268 & 0.398 & 0.211 & 0.355 & 5 & 5 &   2.579 & 5 \\
64794  & 115382   & 8.433 & 0.408 & 0.201 & 0.321 & 3 & 3 &   2.584 & 3 \\
64993  & 115739   & 8.899 & 0.405 & 0.213 & 0.341 & 3 & 3 &   2.588 & 3 \\
65627  & +47 2060 & 9.136 & 0.408 & 0.199 & 0.308 & 3 & 3 &   2.590 & 3 \\
66885  & 119205   & 9.305 & 0.403 & 0.180 & 0.304 & 3 & 3 &   2.562 & 3 \\
70394  & +29 2529 & 9.568 & 0.431 & 0.226 & 0.305 & 4 & 4 &   2.580 & 4 \\
73815  & 133600   & 8.198 & 0.409 & 0.212 & 0.328 & 4 & 5 &   2.600 & 5 \\
74341  & 134902   & 8.861 & 0.416 & 0.214 & 0.335 & 3 & 3 &   2.588 & 3 \\
74389  & 134664   & 7.781 & 0.395 & 0.214 & 0.392 & 3 & 4 &   2.590 & 4 \\
75528  & +47 2225 & 9.785 & 0.427 & 0.192 & 0.262 & 4 & 4 &   2.568 & 4 \\
75923  & 138159   & 9.184 & 0.414 & 0.212 & 0.306 & 2 & 2 &   2.588 & 2 \\
77883  & 142331   & 8.739 & 0.422 & 0.225 & 0.345 & 2 & 2 &   2.594 & 2 \\
77936  & 234267   & 9.334 & 0.412 & 0.151 & 0.292 & 4 & 4 &   2.574 & 4 \\
78028  & +37 2687 & 8.640 & 0.410 & 0.185 & 0.307 & 4 & 4 &   2.590 & 4 \\
78680  & 144270   & 8.198 & 0.404 & 0.182 & 0.299 & 4 & 4 &   2.594 & 4 \\
79186  & 145514   & 8.316 & 0.427 & 0.185 & 0.322 & 4 & 4 &   2.583 & 4 \\
79304  & 145478   & 8.684 & 0.412 & 0.192 & 0.359 & 4 & 4 &   2.594 & 4 \\
79672  & 146233   & 5.494 & 0.400 & 0.219 & 0.356 & 4 & 4 &   2.603 & 4 \\
81512  & +45 2434 & 9.228 & 0.425 & 0.189 & 0.323 & 4 & 4 &   2.581 & 4 \\
85285  & 157691   & 8.388 & 0.406 & 0.164 & 0.312 & 4 & 4 &   2.583 & 4 \\
88194  & 164595   & 7.070 & 0.415 & 0.195 & 0.336 & 4 & 4 &   2.591 & 4 \\
88427  & +35 3136 & 9.331 & 0.418 & 0.179 & 0.314 & 4 & 4 &   2.590 & 4 \\
100963 & 195034   & 7.091 & 0.408 & 0.199 & 0.327 & 4 & 4 &   2.586 & 4 \\
102152 & 197027   & 9.193 & 0.417 & 0.210 & 0.355 & 4 & 4 &   2.595 & 4 \\
103025 & +14 4456 & 8.719 & 0.414 & 0.190 & 0.353 & 4 & 4 &   2.587 & 5 \\
104504 & 201422   & 8.550 & 0.396 & 0.191 & 0.305 & 4 & 4 &   2.589 & 5 \\
108708 & 209096   & 8.943 & 0.415 & 0.217 & 0.353 & 4 & 4 &   2.595 & 4 \\
108996 & 209562   & 8.894 & 0.406 & 0.209 & 0.345 & 4 & 4 &   2.598 & 4 \\
109931 & +24 4563 & 8.941 & 0.423 & 0.205 & 0.350 & 4 & 4 &   2.588 & 4 \\
118159 & 224448   & 9.003 & 0.400 & 0.195 & 0.295 & 3 & 4 &   2.589 & 4 \\
\hline                                 
\end{tabular}
\end{table}
\clearpage
}

\onltab{2}{
\scriptsize
\begin{table}
\caption{Literature photometry (V Johnson and $uvby$-$\beta$) of solar-twin candidates and solar analogs \citep{mer97,hau98}}
\label{tab2}
\centering 
\renewcommand{\footnoterule}{}  
\begin{tabular}{rrrrrrrrrrrrrrrrrrrrrrrrrrrr} 
\hline
\hline 
{HIP number}  & HD number & V & $(b$--$y)$ & $m_1$ & $c_1$ & N$_{V,uvby}$ & $\beta$ & N$_\beta$ \\
\hline
996    & 804     & 8.191 & 0.395 & 0.196 & 0.355 &   2 &        &    \\
11728  & 15632   & 8.035 & 0.414 & 0.221 & 0.335 &   3 &        &    \\
22263  & 30495   & 5.489 & 0.398 & 0.213 & 0.321 &  15 & 2.601  & 11 \\
29525  & 42807   & 6.440 & 0.415 & 0.228 & 0.292 &  13 & 2.593  & 10 \\
30502  & 45346   & 8.660 & 0.411 & 0.202 & 0.331 &   4 &        &    \\
36512  & 59711   & 7.742 & 0.406 & 0.204 & 0.308 &   2 &        &    \\
38228  & 63433   & 6.891 & 0.423 & 0.232 & 0.313 &   1 &        &    \\
41317  & 71334   & 7.812 & 0.412 & 0.210 & 0.327 &   4 &        &    \\
42438  & 72905   & 5.651 & 0.396 & 0.197 & 0.293&    4 &  2.596 & 4  \\
43190  & 75288   & 8.515 & 0.423 & 0.234 & 0.345 &   3 &        &    \\
44713  & 78429   & 7.303 & 0.414 & 0.227 & 0.348 &   3 &  2.600 & 19 \\
44997  & 78660   & 8.345 & 0.410 & 0.216 & 0.321 &   2 &        &    \\
49756  & 88072   & 7.544 & 0.407 & 0.211 & 0.335 &   4 &        &    \\
55409  & 98649   & 8.004 & 0.405 & 0.227 & 0.323 &   1 &  2.588 & 1  \\
55459  & 98618   & 7.658 & 0.411 & 0.198 & 0.347 &   2 &        &    \\
56948  & 101364  & 8.673 & 0.410 & 0.212 & 0.313 &   1 &        &    \\
64150  & 114174  & 6.791 & 0.419 & 0.231 & 0.334 &  15 &  2.592 & 9  \\
77052  & 140538  & 5.869 & 0.424 & 0.231 & 0.334 &   5 &        &    \\
79672  & 146233  & 5.496 & 0.401 & 0.217 & 0.343 &  11 &  2.595 & 3  \\
85042  & 157347  & 6.287 & 0.422 & 0.225 & 0.346 &   4 &        &    \\
100963 & 195034  & 7.090 & 0.408 & 0.204 & 0.319 &   1 &        &    \\
102152 & 197027  & 9.179 & 0.421 & 0.202 & 0.323 &   1 &        &    \\
109110 & 209779  & 7.581 & 0.415 & 0.216 & 0.318 &   2 &        &    \\
\hline
1499   & 1461    & 6.468 & 0.421 & 0.238 & 0.363 &  15 &  2.597 & 13 \\
15457  & 20630   & 4.850 & 0.419 & 0.235 & 0.307 &  54 &  2.595 & 10 \\
53721  & 95128   & 5.037 & 0.391 & 0.202 & 0.343 &  37 &  2.606 & 13 \\
59610  & 106252  & 7.425 & 0.390 & 0.187 & 0.341 &   3 &        &    \\
60081  & 107148  & 8.021 & 0.424 & 0.244 & 0.394 &   2 &        &    \\
62175  & 110869  & 8.008 & 0.412 & 0.215 & 0.358 &   3 &        &    \\
79578  & 145825  & 6.563 & 0.395 & 0.228 & 0.328 &   1 &  2.597 &  1 \\
80337  & 147513  & 5.373 & 0.399 & 0.199 & 0.323 &   5 &  2.609 & 1  \\
96402  & 184768  & 7.556 & 0.426 & 0.214 & 0.343 &   6 &  2.590 &  6 \\
96895  & 186408  & 5.979 & 0.410 & 0.212 & 0.370 &  64 &  2.606 & 39 \\
96901  & 186427  & 6.234 & 0.416 & 0.223 & 0.352 &  63 &  2.597 & 39 \\
113357 & 217014  & 5.456 & 0.416 & 0.232 & 0.371 & 143 &  2.603 & 13 \\
\hline                                 
\end{tabular}
\end{table}
\clearpage
}

\onltab{3}{
\scriptsize
\begin{table}
\caption{Stellar parameters}
\label{tab3}
\centering 
\renewcommand{\footnoterule}{}  
\begin{tabular}{rrrrl} 
\hline
\hline 
{HIP number}  & \teff & log $g$ & [Fe/H] & reference \\
\hline
348      & 5777 & 4.41 & $-$0.13 & R09 \\
996      & 5860 & 4.38 &    0.00 & R09 \\
1499     & 5756 & 4.37 &    0.19 & R09+VF05+LH06+T07+S08 \\
2894     & 5820 & 4.54 & $-$0.03 & R09 \\
4909     & 5836 & 4.44 &    0.02 & R09 \\
5134     & 5779 & 4.49 & $-$0.19 & R09 \\
6407     & 5787 & 4.47 & $-$0.09 & R09 \\
8507     & 5720 & 4.44 & $-$0.08 & R09 \\
8841     & 5676 & 4.50 & $-$0.12 & R09 \\
9349     & 5825 & 4.49 &    0.01 & R09 \\
10710    & 5817 & 4.39 & $-$0.13 & R09 \\
11728    & 5738 & 4.37 &    0.05 & R09+T07 \\
11915    & 5793 & 4.45 & $-$0.05 & R09 \\
14614    & 5803 & 4.47 & $-$0.10 & R09+T07+B10 \\
15457    & 5771 & 4.56 &    0.08 & B10+VF05 \\
18261    & 5891 & 4.44 &    0.00 & R09+T07 \\
22263    & 5826 & 4.54 &    0.00 & B10+VF05 \\ 
25670    & 5755 & 4.38 &    0.07 & R09+T07 \\
28336    & 5713 & 4.53 & $-$0.17 & R09 \\
29525    & 5715 & 4.41 &    0.00 & B10 \\
30502    & 5745 & 4.47 & $-$0.01 & M09 \\
36512    & 5740 & 4.50 & $-$0.09 & M09+T07+S08 \\
38072    & 5839 & 4.53 &    0.06 & R09 \\
38228    & 5693 & 4.52 &    0.01 & R09+VF05 \\ 
41317    & 5724 & 4.46 & $-$0.04 & M09+VF05+S08 \\
42438    & 5864 & 4.46 & $-$0.05 & R09 \\ 
43190    & 5775 & 4.37 &    0.12 & M09 \\
44324    & 5934 & 4.51 & $-$0.02 & R09+T07 \\
44713    & 5784 & 4.36 &    0.10 & B10+VF05+S08 \\
44935    & 5800 & 4.41 &    0.07 & M09 \\
44997    & 5773 & 4.53 &    0.03 & M09+T07 \\
49572    & 5831 & 4.33 &    0.01 & R09 \\
49756    & 5804 & 4.45 &    0.04 & R09+VF05+T07 \\
50826    & 5725 & 4.47 & $-$0.28 & M09 \\
52040    & 5785 & 4.51 & $-$0.09 & R09 \\
52137    & 5842 & 4.56 &    0.07 & R09 \\
53721    & 5916 & 4.48 &    0.03 & B10+VF05 \\
55409    & 5760 & 4.52 & $-$0.01 & M09 \\
55459    & 5838 & 4.42 &    0.04 & R09+VF05+M06+MR07+T07 \\
56948    & 5795 & 4.45 &    0.02 & R09+MR07+T09+M10 \\
59357    & 5810 & 4.45 & $-$0.24 & M09 \\
59610    & 5899 & 4.34 & $-$0.03 & B10+VF05 \\
60081    & 5811 & 4.38 &    0.32 & M09+VF05+S08 \\
60314    & 5874 & 4.52 &    0.11 & R09 \\
60653    & 5725 & 4.38 & $-$0.29 & M09 \\
62175    & 5849 & 4.43 &    0.14 & R09+T07 \\
64150    & 5755 & 4.39 &    0.06 & R09+VF05+T07 \\
64713    & 5815 & 4.52 & $-$0.01 & M09 \\
64794    & 5743 & 4.33 & $-$0.10 & R09 \\
64993    & 5875 & 4.56 &    0.09 & M09 \\
66885    & 5685 & 4.48 & $-$0.38 & M09 \\
73815    & 5803 & 4.34 &    0.02 & MR07+R09 \\
74341    & 5853 & 4.51 &    0.09 & R09 \\
74389    & 5859 & 4.48 &    0.11 & M09+S08 \\
75923    & 5775 & 4.56 & $-$0.02 & M09 \\
77052    & 5697 & 4.54 &    0.04 & B10+VF05 \\
77883    & 5695 & 4.39 &    0.04 & M09 \\
79578    & 5860 & 4.53 &    0.07 & M09+VF05 \\
79672    & 5822 & 4.45 &    0.05 & M09+VF05+M06+MR07+T07+T09+M09 \\
80337    & 5881 & 4.53 &    0.03 & B10+VF05+S08 \\
85042    & 5692 & 4.39 &    0.04 & B10+VF05+S08 \\
85285    & 5730 & 4.43 & $-$0.39 & M09 \\
88194    & 5735 & 4.40 & $-$0.07 & R09+VF05+T07 \\
88427    & 5810 & 4.42 & $-$0.16 & R09 \\
96402    & 5713 & 4.33 & $-$0.03 & B10+T07 \\
96895    & 5808 & 4.33 &    0.10 & R09+VF05+LH06 \\
96901    & 5737 & 4.34 &    0.06 & R09+VF05+LH06+T07 \\
100963   & 5802 & 4.45 &    0.01 & R09+T07+T09 \\
102152   & 5737 & 4.35 & $-$0.01 & R09+M10b \\
104504   & 5836 & 4.50 & $-$0.16 & R09 \\
108708   & 5875 & 4.51 &    0.15 & R09 \\
108996   & 5838 & 4.50 &    0.06 & R09 \\
109110   & 5817 & 4.46 &    0.06 & B10+VF05+T07 \\
109931   & 5739 & 4.29 &    0.04 & R09 \\
113357   & 5803 & 4.38 &    0.22 & R09+VF05+LH06 \\
\hline
\end{tabular}
\end{table}
\clearpage
}

\begin{table}
\caption{Global fits for color = A + B (\teff - 5777) + C (log $g$ - 4.44) + D [Fe/H]}
\label{tabglobal}
\centering 
\renewcommand{\footnoterule}{}  
\begin{tabular}{crrrrrrrrrrrrrrr} 
\hline
\hline 
color &    A   & error  &      B     &    error  &    C     & error   &     D    &  error   & $\sigma$ (fit) & solar colors \\
      &        &        &            &           &          &         &          &          &          &      \\
\hline
$b-y$ & 0.4105 & 0.0005 & -1.2737e-4 & 9.2777e-6 &          &         & 0.049813 & 0.004753 & 0.0046 & 0.4105$\pm$0.0015 \\
$m_1$ & 0.2122 & 0.0006 & -1.1405e-4 & 1.1657e-5 &  0.05051 & 0.00936 & 0.125539 & 0.005877 & 0.0056 & 0.2122$\pm$0.0018 \\
$c_1$ & 0.3319 & 0.0018 &            &           & -0.11223 & 0.02658 & 0.124344 & 0.016006 & 0.0165 & 0.3319$\pm$0.0054 \\
$\beta$& 2.5915 & 0.0008&  4.5183e-5 & 1.5840e-5 & -0.03241 & 0.01192 & 0.034604 & 0.007492 & 0.0061 & 2.5915$\pm$0.0024 \\
\hline 
\end{tabular}
\end{table}

\onltab{5}{
\scriptsize
\begin{table}
\caption{Solar colors and predictions from models atmospheres. 
Our synthetic colors (This work) were computed using the
\cite{bes05} filters, with zero points based on 
both the STIS observed and ATLAS9 synthetic spectrum of Vega}
\label{tab4}
\centering 
\renewcommand{\footnoterule}{}  
\begin{tabular}{lllllll} 
\hline
\hline 
$(b$--$y)$ & $m_1$ & $c_1$ & $\beta$ & reference \\
\hline
\multicolumn{5}{c}{\hbox{\bf Solar colors (this work)}} \\
\hline
 0.4105 ($\pm$0.0015) & 0.2122 ($\pm$0.0018) & 0.3319 ($\pm$0.0054) & 2.5915 ($\pm$0.0024) &    \\
\hline 
\multicolumn{5}{c}{\hbox{\bf Solar spectra}} \\
\hline
0.406,0.390($\pm$0.004), $(b$--$y)$ from $\beta$ &  &   & 2.5955($\pm$0.0024) & \cite{ols76}, using asteroids \\
0.425   & 0.172  & 0.323  & 2.625   & This work, C96 [STIS]\\
0.421   & 0.178  & 0.350  & 2.595   & This work, C96 [ATLAS9]\\
0.424   & 0.188  & 0.306  & 2.632   & This work, N99 [STIS]\\
0.420   & 0.194  & 0.333  & 2.602   & This work, N99 [ATLAS9]\\
0.400   & 0.195  & 0.305  & 2.631   & This work, N99+N03 [STIS]\\
0.396   & 0.201  & 0.332  & 2.601   & This work, N99+N03 [ATLAS9]\\
0.402   & 0.225  & 0.287  & 2.625   & This work, Thuillier (2004) [STIS]\\
0.398   & 0.231  & 0.314  & 2.595   & This work, Thuillier (2004) [ATLAS9]\\
\hline
\multicolumn{5}{c}{\hbox{\bf Kurucz models}} \\
\hline
0.371        & 0.214  &  0.243   &              & \cite{rel78} \\
0.414        &        &          &              & \cite{kur91} \\
0.355, 0.388 & 0.235  & 0.359    & 2.618, 2.660 & \cite{les86} \\
             &        &          & 2.581        & \cite{sma95} \\
0.393(CM), 0.400(noOV), 0.414(OV)  & 0.278 (CM) &  0.339 (CM)  &       & \cite{sma97} \\
0.397(noOVER), 0.410 (OVER) &    &    &   &    \cite{cas97} \\
              &        &         & 2.590 & \cite{cas06} \\
0.410 & 0.208 & 0.276  & 2.617   & This work, ATLAS9 [STIS]\\
0.406 & 0.214 & 0.303  & 2.587   & This work, ATLAS9 [ATLAS9]\\
\hline
\multicolumn{5}{c}{\hbox{\bf MARCS models}} \\
\hline
0.381     & 0.159  & 0.262   &        & \cite{van85} \\
0.383     & 0.258  & 0.325   & 2.589  & O09 \\
0.383     & 0.220  & 0.367   & 2.621  & This work, MARCS 2008 [STIS]\\
0.379     & 0.226  & 0.394   & 2.591  & This work, MARCS 2008 [ATLAS9]\\
\hline 
\multicolumn{5}{c}{\hbox{\bf From empirical calibrations or average of solar analogs}} \\
\hline
0.425($\pm$0.015)  &  &  &  &  \cite{geh81} \\
0.407($\pm$0.010)  &  &  & 2.591($\pm$0.005) & \cite{sax85} \\
0.414($\pm$0.003)  &  &  &  &  \cite{gra92} \\
0.406($\pm$0.004) &  &  &  & \cite{edv93} \\
0.404($\pm$0.005) &  &  &  & \cite{cay96} \\
0.3999            & 0.2090             & 0.323             &   & \cite{cle04} \\
0.394             &  &  &  & RM05b \\
0.403($\pm$0.013) & 0.200($\pm$0.026)  & 0.370($\pm$0.068) &   & \cite{hol06} \\
0.4089($\pm$0.0100) &  &  &  & \cite{cas10} \\
\hline 
\end{tabular}
\end{table}
\clearpage
}

\begin{table}
\caption{$\Delta$ \teff needed to correct the zeropoint of the most recent \teff scales}
\label{tabteff}
\centering 
\renewcommand{\footnoterule}{}  
\begin{tabular}{rll} 
\hline
\hline 
$\Delta$ \teff & reference \\
(K) &  \\
\hline
\multicolumn{2}{c}{\hbox{\bf (b-y)}} \\
\hline
 +137  & \cite{alo96}  \\
 +85   & \cite{gra96}  \\
 +118  & \cite{bla98}  \\
 +68   & \cite{cle04}  \\
 +109  & RM05b   \\
 +48   & \cite{hol07}  \\
 +10  & \cite{cas10}   \\
\hline
\multicolumn{2}{c}{\hbox{\bf $\beta$}} \\
\hline
 +130  & \cite{alo96}   \\
\hline 
\end{tabular}
\end{table}

\begin{table}
\caption{$\Delta$ [Fe/H] needed to correct the zero-point of different $uvby$ metallicity scales}
\label{tabfeh}
\centering 
\renewcommand{\footnoterule}{}  
\begin{tabular}{lllllll} 
\hline
\hline 
$\Delta$ [Fe/H] & $\Delta$ [Fe/H] & reference \\
This work & \'Arnad\'ottir et al. 2010 &  \\
\hline
 +0.06,+0.05  & +0.11$\pm$0.34  & \cite{ols84}  \\
 +0.05        & +0.06$\pm$0.16  & \cite{sch89}  \\
$-$0.04       & +0.00$\pm$0.18  & \cite{hay02}  \\
 +0.07        & +0.05$\pm$0.13  & \cite{mar02}  \\
 +0.07        & +0.06$\pm$0.21  & \cite{mar04}  \\
 +0.04        & +0.04$\pm$0.14  & RM05a \\
 +0.09        & +0.08$\pm$0.16  & \cite{hol07}  \\
 +0.37        & +0.33$\pm$0.30  & O09  \\
\hline 
\end{tabular}
\end{table}


\begin{thebibliography}{}
\bibitem[Allende Prieto et 
al.(2004)]{all04} Allende Prieto, C., Barklem, P.~S., Lambert, D.~L., \& Cunha, K.\ 2004, \aap, 420, 183

\bibitem[Alonso et 
al.(1996)]{alo96} Alonso, A., Arribas, S., \& Martinez-Roger, C.\ 1996, \aap, 313, 873

\bibitem[Alonso et 
al.(1999)]{alo99} Alonso, A., Arribas, S., \& Mart{\'{\i}}nez-Roger, C.\ 1999, \aaps, 140, 261

\bibitem[Arellano Ferro et al.(1990)]{are90} Arellano Ferro, A., Parrao, L., Schuster, W., et al.\ 1990, \aaps, 83, 225

\bibitem[Arellano Ferro 
\& Mantegazza(1996)]{are96} Arellano Ferro, A., \& Mantegazza, L.\ 1996, \aap, 315, 542

\bibitem[\'Arnad\'ottir et al.(2010)]{arn10} \'Arnad\'ottir A. S., Feltzing, S. \& Lundstr\"om, I.\ 2010, \aap, in press

\bibitem[Barry et al.(1978)]{bar78} Barry, D.~C., Cromwell, R.~H., \& Schoolman, S.~A.\ 1978, \apj, 222, 1032

\bibitem[Baumann et al.(2010)]{bau10} Baumann, P., Ram{\'{\i}}rez, I., Mel{\'e}ndez, J., \& Asplund, M. \ 2010, \aap, in press

\bibitem[Bessell et al.(1998)]{bes98} Bessell, M.~S., Castelli, F., \& Plez, B.\ 1998, \aap, 333, 231

\bibitem[Bessell(2005)]{bes05} Bessell, M.~S.\ 2005, \araa, 43, 293 (B2005)

\bibitem[Blackwell \& Lynas-Gray(1998)]{bla98} Blackwell, D.~E., \& Lynas-Gray, A.~E.\ 1998, \aaps, 129, 505

\bibitem[Bohlin(2007)]{bohlin07} Bohlin, R.~C.\ 2007, in: ASP Conference Series, Vol. 364, The Future of Photometric,
Spectrophotometric and Polarimetric Standardization, ed.\ C.\ Sterken, San Francisco:  Astronomical Society of the Pacific, p.~315

\bibitem[Bond \& Neff(1969)]{bon69} Bond, H.~E., \& Neff, J.~S.\ 1969, \apj, 158, 1235 

\bibitem[Bond(1974)]{bon74} Bond, H.~E.\ 1974, \apj, 194, 95

\bibitem[Bond(1980)]{bon80} Bond, H.~E.\ 1980, \apjs, 44, 517

\bibitem[Calamida et al.(2007)]{cal07} Calamida, A., Bono, G., Stetson, P.~B., et al.\ 2007, \apj, 670, 400

\bibitem[Calamida et al.(2009)]{cal09} Calamida, A., Bono, G., Stetson, P.~B., et al.\ 2009, \apj, 706, 1277

\bibitem[Casagrande et al.(2006)]{casagrande06} Casagrande, L., Portinari, L., \& Flynn, C. 2006, \mnras, 373, 13

\bibitem[Casagrande et 
al.(2010)]{cas10} Casagrande, L., Ram{\'{\i}}rez, I., Mel{\'e}ndez, J., Bessell, M., \& Asplund, M.\ 2010, \aap, 512, A54 


\bibitem[Castelli et al.(1997)]{cas97} Castelli, F., Gratton, R.~G., \& Kurucz, R.~L.\ 1997, \aap, 318, 841

\bibitem[Castelli 
\& Kurucz(2004)]{cas04} Castelli, F., \& Kurucz, R.~L.\ 2004, arXiv:astro-ph/0405087

\bibitem[Castelli \& Hubrig(2004)]{cash04} Castelli, F., \& Hubrig, S.\ 2004, \aap, 425, 263 

\bibitem[Castelli(2005)]{cas05} Castelli, F.\ 2005, Memorie 
della Societa Astronomica Italiana Supplement, 8, 25 

\bibitem[Castelli \& Kurucz(2006)]{cas06} Castelli, F., \& Kurucz, R.~L.\ 2006, \aap, 454, 333

\bibitem[Cayrel de Strobel et al.(1981)]{cay81} Cayrel de Strobel, G., Knowles, N., Hernandez, G., \& Bentolila, C.\ 1981, \aap, 94, 1

\bibitem[Cayrel de Strobel(1996)]{cay96} Cayrel de Strobel, G.\ 1996, \aapr, 7, 243

\bibitem[Cayrel de Strobel et 
al.(2001)]{cay01} Cayrel de Strobel, G., Soubiran, C., \& Ralite, N.\ 2001, \aap, 373, 159

\bibitem[Chen et 
al.(2000)]{che00} Chen, Y.~Q., Nissen, P.~E., Zhao, G., Zhang, H.~W., \& Benoni, T.\ 2000, \aaps, 141, 491

\bibitem[Chmielewski(1981)]{chm81} Chmielewski, Y.\ 1981, \aap, 93, 334

\bibitem[Clem et al.(2004)]{cle04} Clem, J.~L., VandenBerg, D.~A., Grundahl, F., \& Bell, R.~A.\ 2004, \aj, 127, 1227

\bibitem[Clements \& Neff(1979)]{cle79} Clements, G.~L., \& Neff, J.~S.\ 1979, \aap, 75, 193

\bibitem[Coelho et 
al.(2005)]{coe05} Coelho, P., Barbuy, B., Mel{\'e}ndez, J., Schiavon, R.~P., \& Castilho, B.~V.\ 2005, \aap, 443, 735 

\bibitem[Colina et al.(1996)]{col96} Colina, L., Bohlin, R.~C., \& Castelli, F.\ 1996, \aj, 112, 307 (C96)

\bibitem[Conti 
\& Deutsch(1966)]{con66} Conti, P.~S., \& Deutsch, A.~J.\ 1966, \apj, 145, 742 

\bibitem[Cox(2000)]{cox00} Cox, A.~N.\ 2000, Allen's Astrophysical Quantities, 4th ed., New York:  AIP Press / Springer

\bibitem[Crawford(1966)]{crawford66} Crawford, D.~L.\ 1966, 
Spectral Classification and Multicolour Photometry, 24, 170 

\bibitem[Crawford \& Barnes(1970)]{cra70} Crawford, D.~L., \& Barnes, J.~V.\ 1970, \aj, 75, 978

\bibitem[Crawford \& Mander(1966)]{cra66} Crawford, D. L., \& Mander, J. 1966, \aj, 71, 114

\bibitem[Crawford(1975)]{cra75} Crawford, D.~L.\ 1975, \aj, 80, 955

\bibitem[Crawford et al.(1972)]{cra72} Crawford, D.~L., Barnes, J.~V., Gibson, J., et al.\ 1972, \aaps, 5, 109

\bibitem[Croft et al.(1972)]{cro72} Croft, S.~K., McNamara, D.~H., \& Feltz, K.~A., Jr.\ 1972, \pasp, 84, 515

\bibitem[Demarque et al.(2004)]{dem04} Demarque, P., Woo, J.-H., Kim, Y.-C., \& Yi, S.~K.\ 2004, \apjs, 155, 667

\bibitem[Edvardsson(2008)]{edv08} Edvardsson, B.\ 2008, Physica Scripta, Volume T 133, pp.~014011

\bibitem[Edvardsson et 
al.(1993)]{edv93} Edvardsson, B., Andersen, J., Gustafsson, B., et al.\ 1993, \aap, 275, 101

\bibitem[Friel 
\& Boesgaard(1992)]{fri92} Friel, E.~D., \& Boesgaard, A.~M.\ 1992, \apj, 387, 170 

\bibitem[Gallou{\"e}tte(1964)]{gal64} Gallou{\"e}tte, L.\ 1964, Annales d'Astrophysique, 27, 423

\bibitem[Gehren(1981)]{geh81} Gehren, T.\ 1981, \aap, 100, 97

\bibitem[Gratton et al.(1996)]{gra96} Gratton, R.~G., Carretta, E., \& Castelli, F.\ 1996, \aap, 314, 191

\bibitem[Gray(1992)]{gra92} Gray, D.~F.\ 1992, \pasp, 104, 1035

\bibitem[Gray(1995)]{gra95} Gray, D.~F.\ 1995, \pasp, 107, 120

\bibitem[Gray(2005)]{gra05} Gray, D.~F.\ 2005, The 
Observation and Analysis of Stellar Photospheres, 3rd Edition. Cambridge, UK:  Cambridge University Press

\bibitem[Gr{\o}nbech et al.(1976)]{gro76} Gr{\o}nbech, B., Olsen, E. H., \& Str\"omgren, B. 1976, A\&AS, 26, 155

\bibitem[Grundahl et al.(2000)]{gru00} Grundahl, F., 
Vandenberg, D.~A., Stetson, P.~B., Andersen, M.~I., 
\& Briley, M.\ 2000, Liege International Astrophysical Colloquia, 35, 503 

\bibitem[Grundahl et al.(2002)]{gru02} Grundahl, F., Briley, M., Nissen, P.~E., \& Feltzing, S.\ 2002, \aap, 385, L14 

\bibitem[Gustafsson \& Nissen(1972)]{gus72} Gustafsson, B., \& Nissen, P.~E.\ 1972, \aap, 19, 261

\bibitem[Gustafsson(2008)]{gus_sun08} Gustafsson, B.\ 2008, Physica Scripta, Volume T 130, pp.~014036

\bibitem[Gustafsson et 
al.(2008)]{gus08} Gustafsson, B., Edvardsson, B., Eriksson, K., et al.\ 2008, \aap, 486, 951

\bibitem[Gustafsson et 
al.(2010)]{gus10} Gustafsson, B., Mel{\'e}ndez, J., Asplund, M. \& Yong, D.\ 2010, ApSS, 328, 185

\bibitem[Hauck \& Kunzli(1996)]{hau96} Hauck, B., \& Kunzli, M.\ 1996, Baltic Astronomy, 5, 303

\bibitem[Hauck \& Mermilliod(1998)]{hau98} Hauck, B., \& Mermilliod, M.\ 1998, \aaps, 129, 431

\bibitem[Hardorp(1978)]{har78} Hardorp, J.\ 1978, \aap, 63, 383

\bibitem[Hardorp(1980a)]{har80a} Hardorp, J.\ 1980a, \aap, 88, 334

\bibitem[Hardorp(1980b)]{har80b} Hardorp, J.\ 1980b, \aap, 91, 221

\bibitem[Haywood(2002)]{hay02} Haywood, M.\ 2002, \mnras, 337, 151

\bibitem[Haywood(2008)]{hay08} Haywood, M.\ 2008, \mnras, 388, 1175


\bibitem[Hilker(2000)]{hil00} Hilker, M.\ 2000, \aap, 355, 994

\bibitem[Holmberg et al.(2006)]{hol06} Holmberg, J., Flynn, C., \& Portinari, L.\ 2006, \mnras, 367, 449

\bibitem[Holmberg et al.(2007)]{hol07} Holmberg, J., Nordstr{\"o}m, B., \& Andersen, J.\ 2007, \aap, 475, 519

\bibitem[Holmberg et al.(2009)]{hol09} Holmberg, J., Nordstr{\"o}m, B., \& Andersen, J.\ 2009, \aap, 501, 941

\bibitem[Israelian et al.(1998)]{isr98} Israelian, G., Garc{\'{\i}}a L{\'o}pez, R.~J., \& Rebolo, R.\ 1998, \apj, 507, 805

\bibitem[Johnson(1962)]{joh62} Johnson, H.~L.\ 1962, \apj, 135, 69

\bibitem[Johnson et al.(1966)]{joh66} Johnson, H. L., Iriarte, B., Mitchell, R. I., \&
Wisniewski, W. Z.\ 1966, Comm. Lunar and Planetary Lab., 4, 99 (Tucson:  Univ. of Arizona)

\bibitem[Joner 
\& Taylor(1995)]{jon95} Joner, M.~D., \& Taylor, B.~J.\ 1995, \pasp, 107, 124 

\bibitem[Jonsell et 
al.(2005)]{jon05} Jonsell, K., Edvardsson, B., Gustafsson, B., et al.\ 2005, \aap, 440, 321

\bibitem[Karata\c{s} \& Schuster(2010)]{kar10} Karata\c{s}, Y., \& Schuster, W.~J. 2010, New Astronomy, 15, 444

\bibitem[Kron(1963)]{kro63} Kron, G.~E.\ 1963, \pasp, 75, 288

\bibitem[Kuiper(1938)]{kui38} Kuiper, G.~P.\ 1938, \apj, 88, 429

\bibitem[Kurucz 
\& Avrett(1981)]{kur81} Kurucz, R.~L., \& Avrett, E.~H.\ 1981, SAO Special Report, 391,  

\bibitem[Kurucz(1991)]{kur91} Kurucz, R.~L.\ 1991, in: Precision Photometry:  Astrophysics of the Galaxy, eds. A.~G.~D.~Philip, A.~R.~Upgren,
and K.~A.~Janes, Schenectady, NY:  Davis Press, p.~27

\bibitem[Kurucz(1993)]{kur93} Kurucz, R.\ 1993, Diatomic 
Molecular Data for Opacity Calculations.~Kurucz CD-ROM No.~15.~Cambridge, 
Mass.: Smithsonian Astrophysical Observatory, 1993., 15,  

\bibitem[Kurucz(1996)]{kur96} Kurucz, R.~L.\ 1996, IAU Symposium, 176, 523 

\bibitem[Kurucz(2005)]{kur05} Kurucz, R.~L.\ 2005, Memorie 
della Societa Astronomica Italiana Supplement, 8, 14 

\bibitem[Lallement et 
al.(2003)]{lal03} Lallement, R., Welsh, B.~Y., Vergely, J.~L., Crifo, F., \& Sfeir, D.\ 2003, \aap, 411, 447 

\bibitem[Lester et al.(1986)]{les86} Lester, J.~B., Gray, R.~O., \& Kurucz, R.~L.\ 1986, \apjs, 61, 509

\bibitem[Lockwood et al.(1992)]{loc92} Lockwood, G.~W., Tueg, H., \& White, N.~M.\ 1992, \apj, 390, 668

\bibitem[Luck \& Heiter(2006)]{luc06} Luck, R.~E., \& Heiter, U.\ 2006, \aj, 131, 3069 (LH06)

\bibitem[Magain(1983)]{mag83} Magain, P.\ 1983, \aap, 122, 225

\bibitem[Malyuto(1994)]{mal94} Malyuto, V.\ 1994, \aaps, 108, 441

\bibitem[Martell \& Laughlin(2002)]{mar02} Martell, S., \& Laughlin, G.\ 2002, \apjl, 577, L45

\bibitem[Martell \& Smith(2004)]{mar04} Martell, S.~L., \& Smith, G.~H.\ 2004, \pasp, 116, 920

\bibitem[Mel{\'e}ndez \& Ram{\'{\i}}rez(2003)]{mel03} Mel{\'e}ndez, J., \& Ram{\'{\i}}rez, I.\ 2003, \aap, 398, 705

\bibitem[Mel\'{e}ndez et al.(2006)]{mel06} Mel\'{e}ndez, J., Dodds-Eden, K. \& Robles, J. A.  2006, \apj, 641, L133 (M06)

\bibitem[Mel\'{e}ndez \& Ram\'{i}rez(2007)]{mel07} Mel\'{e}ndez, J. \& Ram\'{i}rez, I.  2007, \apj, 669, L89 (MR07)

\bibitem[Mel{\'e}ndez et al.(2009)]{mel09} Mel{\'e}ndez, J., Asplund, M., Gustafsson, B., \& Yong, D.\ 2009, \apjl, 704, L66 (M09)

\bibitem[Mel{\'e}ndez et al.(2010a)]{mel10a} Mel{\'e}ndez, J., Casagrande, L., Ram{\'{\i}}rez, I., Asplund, M., \& Schuster, W.~J.\ 2010a, \aap, 515, L3

\bibitem[Mel{\'e}ndez et al.(2010b)]{mel10b} Mel{\'e}ndez, J., Ram\'{i}rez, I., Casagrande, L., et al.\ 2010b, ApSS, 328, 193 (M10b)

\bibitem[Mermilliod et al.(1997)]{mer97} Mermilliod, J.-C., Mermilliod, M., \& Hauck, B.\ 1997, \aaps, 124, 349

\bibitem[Mironov et al.(1998)]{mir98} Mironov, A.~V., Moshkalev, V.~G., \& Kharitonov, A.~V.\ 1998, Astronomy Reports, 42, 799

\bibitem[Mitchell \& Schuster(1985)]{mit85} Mitchell, R.~I., \& Schuster, W.~J.\ 1985, \aj, 90, 2116

\bibitem[Neckel(1986)]{nec86} Neckel, H.\ 1986, \aap, 169, 194
\bibitem[Neckel(1999)]{nec99} Neckel, H.\ 1999, \solphys, 184, 421 (N99)
\bibitem[Neckel(2003)]{nec03} Neckel, H.\ 2003, \solphys, 212, 239 (N03)

\bibitem[Neckel \& Labs(1981)]{nec81} Neckel, H., \& Labs, D.\ 1981, \solphys, 74, 231

\bibitem[Neckel \& Labs(1984)]{nec84} Neckel, H., \& Labs, D.\ 1984, \solphys, 90, 205

\bibitem[Nissen(1981)]{nis81} Nissen, P.~E.\ 1981, \aap, 97, 145 

\bibitem[Nissen \& Schuster(1991)]{nis91} Nissen, P.~E., \& Schuster, W.~J.\ 1991, \aap, 251, 457

\bibitem[Nordstr{\"o}m et al.(2004)]{nor04} Nordstr{\"o}m, B., Mayor, M., Andersen, J., et al.\ 2004, \aap, 418, 989

\bibitem[Olsen(1976)]{ols76} Olsen, E.~H.\ 1976, \aap, 50, 117

\bibitem[Olsen(1983)]{ols83} Olsen, E. H.\ 1983, \aaps, 54, 55

\bibitem[Olsen(1984)]{ols84} Olsen, E.~H.\ 1984, \aaps, 57, 443

\bibitem[{\"O}nehag et al.(2009)]{one09} {\"O}nehag, A., Gustafsson, B., Eriksson, K., \& Edvardsson, B.\ 2009, \aap, 498, 527 (O09)

\bibitem[Parrao et al.(1988)]{par88} Parrao, L., Schuster, W. J., \& Arellano Ferro, A.\
1988, Reporte T\'ecnico \# 52, Instituto de Astronom\'{\i}a, Universidad Nacional
Aut\'onoma de M\'exico

\bibitem[Pasquini et al.(2008)]{pas08} Pasquini, L., Biazzo, K., Bonifacio, P., Randich, S., \& Bedin, L.~R.\ 2008, \aap, 489, 677

\bibitem[Pettit \& Nicholson(1928)]{pet28} Pettit, E., \& Nicholson, S.~B.\ 1928, \apj, 68, 279

\bibitem[Porto de Mello \& da Silva(1997)]{por97} Porto de Mello, G.~F., \& da Silva, L.\ 1997, \apjl, 482, L89

\bibitem[Ram{\'{\i}}rez \& Mel{\'e}ndez(2004)]{ram04} Ram{\'{\i}}rez, I., \& Mel{\'e}ndez, J.\ 2004, \aap, 417, 301

\bibitem[Ram{\'{\i}}rez \& Mel{\'e}ndez(2005a)]{ram05a} Ram{\'{\i}}rez, I., \& Mel{\'e}ndez, J.\ 2005a, \apj, 626, 446 (RM05a)

\bibitem[Ram{\'{\i}}rez \& Mel{\'e}ndez(2005b)]{ram05b} Ram{\'{\i}}rez, I., \& Mel{\'e}ndez, J.\ 2005b, \apj, 626, 465 (RM05b)

\bibitem[Ram{\'{\i}}rez et al.(2009)]{ram09} Ram{\'{\i}}rez, I., Mel{\'e}ndez, J., \& Asplund, M.\ 2009, \aap, 508, L17 (R09)

\bibitem[Ram{\'{\i}}rez et al.(2010)]{ram10} Ram{\'{\i}}rez, I., Asplund, M., 
 Baumann, P., Mel{\'e}ndez, J., \& Bensby, T.\ 2010, \aap, submitted

\bibitem[Reddy et al.(2003)]{red03} Reddy, B.~E., Tomkin, J., Lambert, D.~L., \& Allende Prieto, C.\ 2003, \mnras, 340, 304

\bibitem[Relyea \& Kurucz(1978)]{rel78} Relyea, L.~J., \& Kurucz, R.~L.\ 1978, \apjs, 37, 45

\bibitem[Rieke et al.(2008)]{rie08} Rieke, G.~H., Blaylock, M., Decin, L., et al.\ 2008, \aj, 135, 2245

\bibitem[Saxner \& Hammarback(1985)]{sax85} Saxner, M., \& Hammarback, G.\ 1985, \aap, 151, 372

\bibitem[Sbordone et al.(2004)]{sbo04} Sbordone, L., 
Bonifacio, P., Castelli, F., 
\& Kurucz, R.~L.\ 2004, Memorie della Societa Astronomica Italiana Supplement, 5, 93 

\bibitem[Sbordone et al.(2010)]{sbo10} Sbordone, L., et al.\ 2010, \aap, in press (arXiv:1003.4510)

\bibitem[Schuster(1976)]{sch76} Schuster, W.~J.\ 1976, Rev.\ Mex.\ Astron.\ Astrofis., 1, 327

\bibitem[Schuster \& Nissen(1988)]{sch88} Schuster, W.~J., \& Nissen, P.~E.\ 1988, 
\aaps, 73, 225 (SN)

\bibitem[Schuster \& Nissen(1989)]{sch89} Schuster, W.~J., \& Nissen, P.~E.\ 1989, \aap, 221, 65

\bibitem[Schuster et al.(1993)]{sch93} Schuster, W.~J., Parrao, L., \& 
Contreras Mart{\'{\i}}nez, M.~E.\ 1993, \aaps, 97, 951 (SPC)

\bibitem[Schuster et al.(1996)]{sch96} Schuster, W.~J., Nissen, P.~E., Parrao, L., Beers, T.~C., \& Overgaard, L.~P.\ 1996, \aaps, 117, 317

\bibitem[Schuster \& Parrao(2001)]{sch01} Schuster, W.~J., \& Parrao, L.\ 2001, Rev.~Mex.~Astr.~Astrofis., 37, 187

\bibitem[Schuster et al.(2002)]{sch02} Schuster, W.~J., Parrao, L., \& Guichard, J.\ 2002, The Journal of Astronomical Data, 8, No.~2, 1

\bibitem[Schuster et al.(2004)]{sch04} Schuster, W.~J., Beers, T.~C., Michel, R., Nissen, P.~E., \& Garc{\'{\i}}a, G.\ 2004, \aap, 422, 527

\bibitem[Schuster et al.(2006)]{sch06} Schuster, W.~J., Moitinho, A., M\'arquez, A., Parrao, L., \& Covarrubias, E.\ 2006, \aap, 445, 939

\bibitem[Secchi(1868)]{secchi1868} Secchi, A.\ 1868, Roma : Tip.~Belle Arti, 1868; 68 p.~: tav.~f.~t.; OCCC 4 368 III

\bibitem[Sekiguchi \& Fukugita(2000)]{sek00} Sekiguchi, M., \& Fukugita, M.\ 2000, \aj, 120, 1072

\bibitem[Smalley \& Dworetsky(1995)]{sma95} Smalley, B., \& Dworetsky, M.~M.\ 1995, \aap, 293, 446

\bibitem[Smalley \& Kupka(1997)]{sma97} Smalley, B., \& Kupka, F.\ 1997, \aap, 328, 349

\bibitem[Soderblom \& King(1998)]{sod98} Soderblom, D.~R., \& King, J.~R.\ 1998, in:  Solar Analogs: Characteristics and Optimum Candidates,
ed. J.~C.~Hall, Flagstaff, Arizona:  Lowell Observatory, p.~41

\bibitem[Soubiran \& Triaud(2004)]{sou04} Soubiran, C., \& Triaud, A.\ 2004, \aap, 418, 1089

\bibitem[Sousa et al.(2008)]{sou08} Sousa, S.~G., Santos, N.~C., Mayor, M., et al.\ 2008, \aap, 487, 373 (S08)

\bibitem[Stebbins \& Whitford(1945)]{ste45} Stebbins, J., \& Whitford, A.~E.\ 1945, \apj, 102, 318

\bibitem[Stebbins \& Kron(1957)]{ste57} Stebbins, J., \& Kron, G.~E.\ 1957, \apj, 126, 266

\bibitem[Straizys \& Valiauga(1994)]{str94} Straizys, V., \& Valiauga, G.\ 1994, Baltic Astronomy, 3, 282

\bibitem[Str{\"o}mgren(1963)]{str63} Str{\"o}mgren, B.\ 1963, \qjras, 4, 8 

\bibitem[Str{\"o}mgren(1964)]{str64} Str{\"o}mgren, B.\ 1964, Astrophysica Norvegica, 9, 333

\bibitem[Stromgren et al.(1982)]{str82} Stromgren, B., Gustafsson, B., \& Olsen, E.~H.\ 1982, \pasp, 94, 5 

\bibitem[Takeda et al.(2007)]{tak07} Takeda, Y., Kawanomoto, S., Honda, S., Ando, H., \& Sakurai, T.\ 2007, \aap, 468, 663 (T07)

\bibitem[Takeda \& Tajitsu(2009)]{tak09} Takeda, Y., \& Tajitsu, A.\ 2009, \pasj, 61, 471 (T09)

\bibitem[Taylor(1984)]{tay84} Taylor, B.~J.\ 1984, \apjs, 54, 167

\bibitem[Taylor(1994)]{tay94} Taylor, B.~J.\ 1994, \pasp, 106, 444

\bibitem[Thuillier et al.(2004)]{thui04} Thuillier, G., Floyd, L., Woods, T.~N., et al.\ 2004, Adv.\ Space Res., 34, 256

\bibitem[Tueg(1982)]{tue82} Tueg, H.\ 1982, \aap, 105, 395

\bibitem[Tueg \& Schmidt-Kaler(1982)]{tue_sch82} Tueg, H., \& Schmidt-Kaler, T.\ 1982, \aap, 105, 400

\bibitem[Twarog et al.(2002)]{twa02} Twarog, B.~A., Anthony-Twarog, B.~J., \& Tanner, D.\ 2002, \aj, 123, 2715

\bibitem[Twarog et al.(2007)]{twa07} Twarog, B. A., Vargas, L. C., Anthony-Twarog, B. J.\ 2007, \aj, 134, 1777

\bibitem[Valenti \& Fischer(2005)]{val05} Valenti, J.~A., \& Fischer, D.~A.\ 2005, \apjs, 159, 141 (VF05)

\bibitem[van den Bergh(1965)]{van65} van den Bergh, S.\ 1965, \jrasc, 59, 253

\bibitem[Vandenberg \& Bell(1985)]{van85} Vandenberg, D.~A., \& Bell, R.~A.\ 1985, \apjs, 58, 561

\bibitem[Vandenberg \& Poll(1989)]{van89} Vandenberg, D.~A., \& Poll, H.~E.\ 1989, \aj, 98, 1451

\bibitem[Welsh et 
al.(2010)]{wel10} Welsh, B.~Y., Lallement, R., Vergely, J.-L., \& Raimond, S.\ 2010, \aap, 510, A54 

\bibitem[Wielen et al.(1996)]{wie96} Wielen, R., Fuchs, B., \& Dettbarn, C.\ 1996, \aap, 314, 438

\bibitem[Woods et al.(1996)]{woo96} Woods, T.~N., Prinz, D.~K., Rottman, G.~J., et al.\ 1996, J. Geophys. Res., 101, 9541

\bibitem[Yong et al.(2008)]{yon08} Yong, D., Grundahl, F., 
Johnson, J.~A., \& Asplund, M.\ 2008, \apj, 684, 1159 

\bibitem[Zacs et al.(1998)]{zac98} Zacs, L., Nissen, P.~E., \& Schuster, W.~J.\ 1998, \aap, 337, 216 

\end{thebibliography}
\end{document}